\documentclass[a4paper,fleqn,usenatbib]{mnras}

\usepackage[T1]{fontenc}
\usepackage{ae,aecompl}
\usepackage{graphicx}
\usepackage{epsfig}
\usepackage{amsmath}
\usepackage{amssymb}

\def\beq{\begin{equation}}
\def\eeq{\end{equation}}
\def\beqar{\begin{eqnarray}}
\def\eeqar{\end{eqnarray}}

\def\avg#1{\langle #1 \rangle}

\def\ecrsn{\epsilon_{\rm cr}}

\def\ycr{y^{\rm cr}}

\title{Are starburst galaxies proton calorimeters?}

\author[X. Wang and B. D. Fields]{Xilu Wang,$^{1}$\thanks{E-mail: xwang107@illinois.edu}
and Brian D. Fields,$^{1,2}$
\\
$^{1}$Department of Astronomy, University of Illinois, Urbana, IL 61801, USA\\
$^{2}$Department of Physics, University of Illinois, Urbana, IL 61801, USA\\
}

\begin{document}
\label{firstpage}
\pagerange{\pageref{firstpage}--\pageref{lastpage}}
\maketitle

\begin{abstract}
Several starburst galaxies have been observed in the GeV and TeV bands.
In these dense environments,
gamma-ray emission should be dominated by cosmic-ray interactions 
with the interstellar medium ($p_{\rm cr}p_{\rm ism} \to \pi^{0} \to \gamma\gamma$). 
Indeed, starbursts may
act as proton ``calorimeters''
where a substantial fraction of
cosmic-ray energy input is emitted in gamma rays.
Here we build a one-zone,  ``thick-target'' model implementing calorimetry and
placing a firm upper bound on gamma-ray emission from cosmic-ray interactions.
The model assumes that cosmic rays are accelerated by supernovae (SNe), and all suffer nuclear interactions rather than escape. 
Our model has only two free parameters: the cosmic-ray proton acceleration energy per supernova $\epsilon_{\rm cr}$,
and the proton injection spectral index $s$.
We calculate the pionic gamma-ray emission from 10 MeV to 10 TeV, and
 derive thick-target parameters for six galaxies with {\em Fermi}, {\em H.E.S.S.}, and/or {\em VERITAS} data. 
Our model provides good fits for the M82 and NGC 253, 
and yields $\epsilon_{\rm cr}$ and $s$ values 
suggesting that supernova cosmic-ray acceleration
is similar
in starbursts and in our Galaxy.
We find that these starbursts are indeed nearly if not fully proton calorimeters.
For NGC 4945 and NGC 1068, the models are consistent with calorimetry but
are less well-constrained due to the lack of TeV data. 
However, the Circinus galaxy and the ultraluminous infrared galaxy Arp 220 exceed our pionic upper-limit; possible explanations are discussed. 
\end{abstract}


\section{Introduction}
\label{sec:Introduction}

Cosmic rays (CRs) are accelerated by 
supernovae \citep[e.g.,][]{Zwicky, Ginzburg & Syrovatskii 1964, Ackermann et al. 2013},
and thus cosmic-ray production is an inevitable consequence of star formation. 
As CRs propagate in the interstellar medium (ISM),
inelastic collisions between CR and interstellar nuclei--both dominantly protons--lead
to gamma-ray production via $\pi^{0}$ decay: $p_{\rm cr}p_{\rm ism} \to \pi^{0} \to \gamma \gamma$ \citep{Stecker 1971, Dermer 1986}. 
This process occurs not only in the Milky Way, but also in other star-forming galaxies \citep[e.g.,][]{Dermer 1986, Strong 1976, Lichti 1978, Pavlidou & Fields 2001, Stecker & Venters 2011, Abdo et al. 2009, Fields et al. 2010, Strong et al. 2010}. 
Compared with normal star-forming galaxies like Milky Way, starbursts and ultraluminous infrared galaxies (ULIRGs, the very extreme starbursts) have exceptionally high star-formation rates and harbor regions of very dense gas.
Thus cosmic rays accelerated in starbursts are expected to be lost due to interaction rather than escape, whereas normal star-forming galaxies are
in the opposite regime.
In the limit where all of the CR nuclei interact with ISM rather than escape,
a large fraction of initial proton energy is emitted as gamma rays, making
such a galaxy a ``proton calorimeter''  \citep[e.g.,][]{Pohl1993,Pohl1994,Lacki et al. 2011,Abramowski et al. 2012}.\footnote{
A closely analogous concept is cosmic-ray electron calorimetry,
as suggested observationally by, e.g., the far infrared--radio correlation \citep[e.g.,][]{voelk}.
} 
This situation has the maximum efficiency to convert supernova blast energy into gamma rays. 
Therefore the starbursts galaxies were anticipated to be detected as gamma-ray sources \citep[e.g.,][]{Paglione et al. 1996,Blom et al. 1999,Domingo & Torres 2005,Persic et al. 2008,de Cea del Pozo et al. 2009,Rephaeli et al. 2010}.

{\em Fermi} LAT is the first gamma-ray telescope to observe the starburst galaxies, and is also the first one to study external star-forming galaxies as a population.
Three of the {\em Fermi} detections are normal star-forming galaxies: 
the Large Magellanic Cloud \citep[LMC][]{Abdo et al. 2010a}, 
the Small Magellanic Cloud \citep[SMC][]{Abdo et al. 2010b}, and M31 \citep{Abdo et al. 2010c}.
Five additional {\em Fermi} detections are starburst galaxies: M82 and NGC 253 \citep{Abdo et al. 2010d}, 
NGC 4945 and NGC 1068 \citep{Nolan et al. 2012}, as well as the Circinus galaxy \citep{Hayashida et al. 2013}. 
The two nearest and brightest starbursts, M82 and NGC 253, are also detected at TeV energies by \textit{VERITAS} \citep{VERITAS Collaboration et al. 2009} and \textit{H.E.S.S.} \citep{Acero et al. 2009, Abramowski et al. 2012}, respectively. \citet{Peng 2016} and \citet{Griffin 2016} recently reported {\em Fermi} detections
of the ULIRG Arp 220. 
Star-forming galaxies represent a new gamma-ray source class, and offer unique insight into the global behavior
of cosmic rays over a wide range of galaxy types and star-formation rates.

Various models have been built for starbursts to study the multi-frequency emissions from radio to $\gamma$-rays, considering both hadronic and leptonic processed (e.g., synchrotron radiation, inverse Compton scattering (IC), pion production). For example, \citet{Blom et al. 1999}, \citet{Persic et al. 2008}, \citet{de Cea del Pozo et al. 2009}, \citet{Lacki et al. 2011}, \citet{Lacki et al. 2014}, \citet{PA2012}, \citet{YH2013} give their predictions of gamma-radiation from M82, while NGC 253 are anticipated to be observed in GeV-TeV range by \citet{Paglione et al. 1996}, \citet{Domingo & Torres 2005}, \citet{Rephaeli et al. 2010}, \citet{Lacki et al. 2011}; \citet{Lacki et al. 2014}, \citet{PA2012}, \citet{YH2014}. Recent observations and current theoretical models of starbursts are also reviewed by \citet{Ohm 2016}.
Many-but not all-of these models predict that hadronic processes dominate above a few GeV. In this paper, our aim is to calculate
self-consistently the pionic emission from starbursts in a closed box, and to use starburst data to test this calorimetric scenario. By construction, our more focused model is economical and thus easy to test: it contains only two parameters, the cosmic-ray acceleration energy per supernova $\epsilon_{\rm cr}$, and the cosmic-ray injection index $s$. Some early results from our calculations were summarized in \citet{Wang & Fields 2014} and \citet{Wang:2015omn}. 

In this paper, we define a {\em proton calorimeter} to be a system in which cosmic-ray pionic losses dominate over other losses including escape, advection, and diffusion.  Such a system is in the ``thick-target'' regime of cosmic-ray propagation, and a substantial fraction of the energy injected into cosmic-ray protons energy is ultimately emitted as pionic gamma-ray photons. The calorimetric efficiency (eq.~\ref{eq:efficiency}) is a measure of gamma-ray energy output to the cosmic-ray energy input.

The next section shows the assumptions, important expressions and physics of our thick-target model. \S~\ref{sec:Model Results} presented the results calculated from our model when applying to five observed starbursts galaxies and the ULIRG Arp 220. In \S~\ref{sec:Discussion and Conclusions}, further discussions and conclusions are given.

\section{The Thick-Target/Calorimetric Model}
\label{sec:The Thick-Target/Calorimetric Model}

To calculate the hadronic gamma-ray output in out model, we first
characterize the cosmic-ray sources and their thick-target propagation.
We then use the propagated cosmic-ray flux to arrive at
hadronic gamma-ray emission. The calculation in this session adopts GeV as the energy unit.

\subsection{Model Assumptions}
\label{subsec:Assumptions}

We describe the production and propagation of cosmic rays
in a one-zone, thick-target ``closed-box'' model. The physical processes in our model are CR ion acceleration by SNe, followed by pion production through the interaction between the CRs and the ISM. The resulting neutral pion decay is responsible for the existing gamma-rays. The basic assumptions are:
\begin{enumerate}
\item
  cosmic-rays and ISM gas are both spatially homogeneous;

\item
  cosmic rays are accelerated by supernovae (SNe) with acceleration energy per SN $\epsilon_{\rm cr}$;

\item
  the injected cosmic-ray/proton spectrum is a power law in momentum, of spectral index $s$ in GeV and TeV energy range;

\item
  all the cosmic rays will interact with ISM, i.e. the escape rate of protons is zero, {\em advection and diffusion loss are also ignored here}; and

  \item
    among the gamma-ray production mechanisms, pion production and decay dominates.
\end{enumerate}

Our thick-target model places a firm upper-limit on the hadronic (pionic)
gamma-ray emissions from starbursts, by including only losses due to particle interactions (collisions and scattering). Other work
has argued that in starburst regions, the dense gas, high supernova density and relatively hard gamma ray spectrum point to
diffusion and advection losses being subdominant \citep[e.g.,][]{Lacki2010, Lacki & Thompson 2013,Torres2012}.
We concur, and in Appendix \ref{App:AppendixB}, we show that in starbursts, the interaction time
is much sorter than the diffusion and advection times.   We thus omit these effects in our model for an upper-limit calculation.
If the advection and diffusion losses were included, the actual calorimetric efficiencies are reduced, which may explain the difference in the starbursts' calorimetric efficiencies obtained with our thick-target model in \S~\ref{sec:Model Results} (due to the different values of $\tau_{\rm diff}$ and $\tau_{\rm adv}$ in each starburst).
We also neglect reacceleration of cosmic rays \citep{strong07} inside starbursts, which merits a study in its own right.
\footnote{Note that  reacceleration due to supernova shocks would be an additional way to transfer supernova blast
  energy to cosmic rays, and so would amount to a component of calorimetry. If other shock contribute to
  reacceleration, it would give an apparent boost to the calorimetric efficiency.}

The thick-target hadronic model presented here neglects primary electron effects (bremsstrahlung radiation, inverse Compton) and secondary electron effects in gamma-ray emission. This assumption is consistent with {\em NuSTAR}'s
upper limits on NGC 253 in the 7-20 keV band, which disfavor leptonic processes dominating in the GeV and TeV energy range \citep{Nustar}. In addition, \citet{Strong et al. 2010} found pionic emission dominates over both primary and secondary electron emission by factors $> 2$ among the total Galactic luminosity in GeV range. This implies
that in starbursts where cosmic ray proton losses dominate over escape, pionic emission should be even
more dominant over leptonic. 
By assuming the CR protons lose energy continuously through the propagation inside the starbursts, the effect of secondary recoil protons (the ISM protons after $pp$ collisions) appears only via the elastic scattering energy loss term, and not as a proton source term.  While these effects are not large,
they would only boost the gamma-ray production and lead to an even tighter limit to the gamma-ray emission. 

We also ignore the effect of intergalactic absorption of the high-energy gamma rays via photon-pair production ($\gamma\gamma \to e^{+}e^{-}$) in collision with background starlight emission \citep[e.g.,][]{Salamon & Stecker 1998, Stecker et al. 2012} and in collision with the infrared field of the starbursts \citep{Lacki & Thompson 2013}. The former effect will bring a steepening of the gamma-ray spectrum at high energy, but this effect is very small for the starbursts we study, which are
all very nearby. The later effect can be substantial for gamma-ray energies above a few TeV, 
but is negligible in the GeV energy range that is our focus.

\subsection{Cosmic-Ray Source and Propagation}
\label{subsec:Cosmic-Ray Source and Propagation}

The equations for cosmic-ray transport 
\citep[e.g.,][]{Longair 1981, strong07,Mene1971,Fields 1994}
can be written as
\beq
\label{eq:prop}
\partial_t N_E = \partial_E (b_E N_E) - \frac{1}{\tau_E} N_E + q_E \
+ \mbox{advection} + \mbox{diffusion} \ .
\eeq
Here and throughout, $E$ denotes {\em kinetic} energy per nucleon,
and $N_E \ dE$ is the number density of cosmic rays with kinetic energy $\in (E,E+dE)$.
The cosmic-ray number flux density
is thus $\phi(E) = v_E N_E$, with $v_E$ the
velocity at $E$.
In eq.~(\ref{eq:prop}), $\tau_E$ is the lifetime of cosmic ray against escape, $q_E$ is the injected cosmic ray spectrum, $b_E=-dE/dt$ is the rate of
energy loss (per nucleon).

We now drastically simplify the problem, adopting the closed-box,
thick-target, steady-state limit corresponding to the discussion
in \S\ref{subsec:Assumptions}.
That is, we focus on a single uniform zone, in which
cosmic rays are accelerated and then propagate until lost
due to their interactions, and in which acceleration and losses
are driven to an equilibrium $\partial_t N_E=0$
over the energy loss timescale $\tau_{\rm loss}=\int dE/b \sim E/b$. 
We thus neglect escape, so that $1/\tau_E=0$,
and spatial uniformity implies that the gradient-driven
advection and diffusion terms are zero.
 
The closed-box, steady-state solution to eq.~(\ref{eq:prop})
gives a proton flux density
\beq
\label{eq:spectrum_p}
\phi_{\rm p} \left(E\right)= v N_E = \frac{v_{\rm p}}{b\left(n_{\rm gas},E\right)} \int_{E}^{\infty} dE' q_{\rm E}\left(E'\right)
\eeq
We see that in this simple model, the cosmic-ray flux depends
only the cosmic ray source function $q_E$ and energy loss rate $b$. 
  
Since cosmic rays accelerated by the supernovae in our model, energy conservation implies
\beq
\label{eq:injectedluminosity}
L_{\rm cr}= \frac{dE_{\rm cr}}{dt}=E_{\rm sn}f_{\rm cr} R_{\rm sn}=\epsilon_{\rm cr} R_{\rm sn}=V \int_{E_{\rm min}}^{\infty} E \frac{dq}{dp} dp
\eeq
where $L_{\rm cr}$ is the injected cosmic ray luminosity, $V$ is the volume of the galaxy where cosmic rays
are produced, $E_{\rm min}$ is the minimum kinetic energy of injected protons that can be accelerated. $E_{\rm sn}$ is the total baryonic energy released by one SN explosion.  Some fraction $f_{\rm cr}$ of this explosion energy goes to accelerate cosmic rays, and
this leads to the other free parameter in our model: $\epsilon_{\rm cr}=E_{\rm sn}f_{\rm cr}$the cosmic-ray proton acceleration energy per supernova.
$R_{\rm sn}$ is the SN rate, which can be converted from the star formation rate (SFR) $\psi$ by $R_{\rm sn}/\psi \sim 0.00914 M_{\odot}^{-1}$ \citep{Lien & Fields 2009}. 

Following the simplest (i.e., test particle) expectations of diffusive shock acceleration \citep[e.g.,][]{Krymskii1977, Bell1978, Blandford1978} 
we assume the injected cosmic ray spectrum (emissivity) is
a power law in momentum:
\beq
\label{eq:cr_spectrum}
q_{\rm p}= \frac{dN}{dV dt dp}=\frac{dq}{dp}=\frac{q_0}{I} p^{-s}
\eeq
where $q_0=L_{\rm cr}/V$ is the cosmic ray luminosity density, $I$ is the normalization factor, $s$, the proton spectral injection index, is a free parameter in the model ($>2.0$).
See \S\ref{subsec: Index and SNR} for more discussion of this assumption.

Finally we can get the accelerated proton spectrum:
\beq
\label{eq:protonspectrum}
\phi_{\rm p} \left(E\right) = \frac{q_0 v_{\rm p}} {I b(n_{\rm gas}, E)}\frac {(p_{\rm p})^{1-s}} {s-1}
\eeq
where $p_{\rm p}$ is proton's momentum,  and $I$ is a number that is determined by $E_{\rm min}$: $I=I\left(E_{\rm min}\right)=\int_{E_{\rm min}}^{\infty} E (E+m_{\rm p}) p_{\rm p}^{-s-1} dE$, $m_{\rm p}$ is the mass of proton.

\subsection{Pionic Emission From Thick-Target Galaxies: the Calorimetric Model }
\label{subsec:Pionic Emission From Individual Starburst Galaxies}

Our notation and approach follows that of
\citet{Dermer 1986}.
From the accelerated proton spectrum, we can get the pionic spectrum (in the lab frame) through the interaction $p_{\rm cr}p_{\rm ism} \to \pi^{0} \to \gamma\gamma$: 
\beq
\label{eq:pionspectrum}
\frac{dq_{\pi}(E_{\pi})}{dE_{\pi}} = n_{\rm gas}\int_{E_{\rm p}^{\rm threshold}}^{\infty} dE_{\rm p} \phi_{\rm p}(E_{\rm p}) \frac{d\sigma_{\pi}(E_{\pi},E_{\rm p})}{dE_{\pi}}
\eeq
In turn, the gamma-ray spectrum is 
\beqar
\label{eq:gammaspectrum}
& &\frac{dq_{\gamma}(E_{\gamma})}{dE_{\gamma}}[{\rm photons/(cm^3-s-GeV)}] 
\nonumber\\
&=&2\int_{E_{\gamma}+(m_{\pi}^2/4E_{\gamma})-m_{\pi}}^{\infty} dE_{\pi} \frac{dq_{\pi}(E_{\pi})/dE_{\pi}}{((E_{\pi}+m_{\pi})^2-m_{\pi}^2)^{1/2}}
\eeqar
where $E_p^{\rm threshold}(E_{\pi})$ is the threshold proton kinetic energy that can produce a pion with energy $E_\pi$, and $m_{\pi}$ is the $\pi^0$ mass.
The differential cross section $d \sigma_{\pi}(E_p,E_{\pi})/dE_{\pi}$  for the production of a $\pi^0$ with energy $E_{\pi}$ can be written as $d \sigma_{\pi}(E_p,E_{\pi}) / dE_{\pi}=\avg{\zeta \sigma_{\pi}(E_{\rm p})} dN(E_{\rm p},E_{\pi})/dE_{\pi}$.  Here $\avg{\zeta \sigma_{\pi}(E_{\rm p})}$ is the inclusive cross section for the reaction $p_{\rm cr}p_{\rm ism} \to \pi^{0} \to \gamma\gamma$.

Our model self-consistently calculates the inelastic energy loss from cross-section $\avg{\zeta \sigma_{\pi}(T_{\rm p})}$. We use the \citet{Dermer 1986}
for the inclusive cross-section $\avg{\zeta \sigma_{\pi}(E_{\rm p})}$, and thus we can get the inelastic energy loss rate consistently (assuming the loss is approximated to be continuous): 
\beq
\label{eq:inelasticb}
b_{\rm inelastic}= 3 n_{\rm gas} v_{\rm p}  \avg{\zeta \sigma_{\pi}(E_{\rm p})}  \int_{0}^{E_{\rm p}} E_{\pi} dN(E_p,E_{\pi})/dE_{\pi}dE_{\pi} . \
\eeq
The crucial factor of 3 here comes from assuming the inclusive cross sections for $p_{\rm cr}p_{\rm ism} \to \pi^{\pm} +{\rm anything}$ are the same as $\pi^0$, i.e., the production rates for $(\pi^-,\pi^0,\pi^+)$ are approximately the same. This factor  of 3 has a direct impact on the gamma-ray production efficiency:
the gamma energy output per energy into CRs above pionic threshold would
be 1/3 if the inelastic losses were the only ones.

We also include the energy loss contributions due to nuclear elastic scattering \citep{Gould 1982} and ionization \citep{Ginzburg & Syrovatskii 1964}: $b(n_{\rm gas},E)=b_{\rm inelastic}+b_{\rm elastic}+b_{\rm ioniz}$, with rates given in the Appendix \ref{App:AppendixA}.  These two terms also affect the gamma-ray production efficiency:  ionization loss is only important at low energies, but the elastic scattering is important at all energies and in general is comparable to inelastic. Thus elastic losses are the more important to lower the CR efficiency.

The function $dN(E_{\rm p},E_{\pi})/dT_{\pi}$ encodes the distribution
of pion energies at each proton energy.
We adopt \citet{Dermer 1986}'s approach, combining Stecker's isobaric model \citep[model S,][]{Stecker1970} together with Stephens and Badhwar's scaling model \citep[model SB,][]{SB1981}: for $E_{\rm p}<3 {\rm GeV}$, model S is used; while model SB is adopted for $E_{\rm p}>7 {\rm GeV}$; for $3 {\rm GeV}<E_{\rm p}<7 {\rm GeV}$, model S and model SB is linearly connected to be used.

Collecting these results gives the emissivity 
\beqar
\label{eq:gammarayspectrum}
\frac{dq_{\gamma}}{dE_{\gamma}} & = & \frac {dN_{\gamma}}{dVdE_{\gamma}dt}
\nonumber\\
& = & \frac{\epsilon_{\rm cr} R_{\rm sn}}{V I \left(E_{\rm min} \right)} I_{0}(E_{\gamma},s)
\eeqar
where $I_{0}(E_\gamma,s)$ is a dimensionless integration:
\beqar
\label{eq:I0}
I_{0}(E_\gamma,s) & = & \int_{E_{\gamma}+{m_{\pi}}^{2}/\left(4E_{\gamma}\right)-m_{\pi}}^{\infty} \frac{2dE_{\pi}}{p_{\pi}}\int_{E^{\rm threshold}_p}^{\infty}dE_{\rm p}
\nonumber\\
 & & \times \frac{\sigma_{\pi}\left(E_{\rm p}\right)}{b(E_{\rm p})}\frac{dN\left(E_{\pi},T_{\rm p}\right)}{d\left(E_{\pi}\right)} 
n_{\rm gas} v_{\rm p}\frac{p_{\rm p}^{1-s}}{s-1}
\eeqar

Notice that the energy loss rate scales with gas density:  
$b \propto n_{\rm gas}$ (see eqs.~\ref{eq:inelasticb}, \ref{eq:be}, and \ref{eq:bi}).
This exactly cancels the gas density in the numerator of eq.~(\ref{eq:I0}),
and thus {\em the gamma-ray emission is independent of the gas density for the thick-target model.}
This is characteristic of calorimetry.  Note further that the ratio $b/n_{\rm gas}$ 
depends only on the cross sections in the loss interactions.
This means that $I_0$ and thus the gamma-ray emission depends only on the ratio
of cross sections (inelastic pion production to total losses).

To account for the contribution from particle interactions involving nuclei with atomic weights $A>1$ in both CRs and ISM, a nuclear enhancement factor of ${\cal A}=0.59$ is included in the calculation. In the case of calorimetry, Appendix \ref{App:AppendixC} shows that the ``nuclear enhancement'' ${\cal A} = 1/\langle A \rangle$ and so ${\cal A} < 1$, this arises because additional nuclei species must share a fixed CR injection energy budget.

Let $d$ to be the distance of the source, $\gamma$-ray flux can be expressed as:
\beq
\label{eq:gammaflux1}
E_{\gamma}^{2}F_{\gamma} = E_{\gamma}^{2}\frac {dN_{\gamma}}{dE_{\gamma}dAdt} = {\cal A}\frac{\epsilon_{\rm cr} R_{\rm sn}}{ I(E_{\rm min})}  E_{\gamma}^{2} \frac{1}{4\pi{d}^2} I_{0}(E_{\gamma},s)
\eeq
and the gamma-ray energy luminosity from the galaxy is:
\beqar
\label{eq:gammalum}
L_{\gamma} & = & \frac {dE_{\gamma}}{dt} = \int E_{\gamma}\frac{dq_{\gamma}}{dE_{\gamma}} dE_{\gamma}dV 
\nonumber\\
& = & {\cal A} \frac{\epsilon_{\rm cr} R_{\rm sn}}{ I(E_{\rm min})}  \int dE_{\gamma} E_{\gamma} I_{0}(E_{\gamma},s)
\eeqar
Note that the volume integration in our one-zone model
cancels the factor in the emissivity $q_\gamma$ (eq.~\ref{eq:gammarayspectrum}), leading to the final result that is independent of volume.
 We see therefore that in our calorimetric limit,
the ratio $L_\gamma/R_{\rm SN}$ depends on the supernova acceleration parameters
$\epsilon_{\rm cr}$ and $s$, as well as $I_0$ that depends only
on cross sections. It is {\em independent} of the gas density, mass, and volume in this calorimetric model.

The luminosity $L_{\gamma} \propto R_{\rm sn}$, and while the SN rate
is usually not measured directly, its is proportional to the star-formation rate of a galaxy.  Therefore we can get
\beqar
\label{eq:gammalumSFR}
\frac{L_{\gamma}}{\psi}  =  \frac{0.00914 M_{\odot}^{-1} {\cal A}\epsilon_{\rm cr} }{ I(E_{\rm min})}  \int dE_{\gamma} E_{\gamma} I_{0}(E_{\gamma},s)
\eeqar
that is a constant only depend on CR proton spectral index $s$ in a calorimetric limit ($\epsilon_{\rm cr}=0.3\ {\rm foe}$, with $1\ {\rm foe} \equiv 10^{51} {\rm erg} \equiv 1\ {\rm Bethe}$). $L_{\gamma}/\psi$ is observable, so it can be used to investigate cosmic-ray properties in a calorimetric system.

We can see that, our model's gamma-radiation results only depend on two parameters: cosmic-ray proton acceleration energy per supernova $\epsilon_{\rm cr}$ (direct proportionality) and the proton injection spectral index $s$. We only need to vary the two parameters $\epsilon_{\rm cr}$ and $s$ to find the best fit to the model (\S \ref{subsec: Index and SNR}).
An order of magnitude calculation of our model in Appendix \ref{sec:Order-Of-Magnitude Estimation} helps to give intuition for the final results and frame key physical issues.

\subsection{Projectile CR Proton Index And Supernova Acceleration Energy/Efficiency}
\label{subsec: Index and SNR}

In our model, each supernova accelerates cosmic rays, which are 
lost via interactions with interstellar gas, and the $\pi^0$
from these interactions give rise to gamma rays.  Thus the gamma-ray
output ultimately depends on the CR properties of the supernova
sources:  the proton injection index $s$ and CR acceleration energy per SN are the only two parameters our model.
Milky Way supernova remnant (SNR) gamma-ray data together with supernova acceleration theories can give both observational and theoretical insight into the parameters we have derived for starbursts in the previous section.

Diffusive shock acceleration naturally yields a relativistic electron and ion spectra that are each power laws in momentum, in the test-particle limit that neglects feedback from the accelerated cosmic rays onto the shock \citep[e.g.,][]{Krymskii1977, Bell1978, Blandford1978}.
Although the resulting non-linear correction to diffusive shock acceleration results in a concave proton spectrum with a steeper spectrum index at high energy \citep[e.g.,][]{Blasi 2016, Jones 2013, Ellison 2014}, the concavity is expected to be rather mild for a SN with particle acceleration efficiency to be at the order of $\sim10\ \rm{percent}$ \citep{Blasi 2016}. 

For a strong shock in monatomic gas, diffusive shock acceleration gives $s \rightarrow 2.0$. 
In GeV and TeV energy range, the combination of observed CR flux at Earth ($\propto E^{-2.75}$) and galactic CR transportation models \citep[e.g.,][]{Strong & Moskalenko 1998, Evoli et al. 2008, Blasi & Amato 2012} implies the index $s$ to be $2.2-2.4$ \citep{Caprioli 2012}. Other theories give different values of the source proton index value in SNR, for example, \citet{Dermer et al. 2013} gives 2.5 below 6.5 GeV and 2.8 above, for the interstellar cosmic-ray proton index; 
\citet{Morlino & Caprioli 2012}'s model for SNR Tycho gives $s=2.2$. 
Gamma-ray emission from SNRs probes $s$ directly (if pions dominate), 
and available measurements give $s$ spanning a considerable range.
 \textit{Fermi} LAT measurement of Galactic SNRs  give $s = 1.53$ to 3.58 with the weighted average to be 2.39, while the spread of the index is about 1 \citep{Acero et al. 2016}.
 Because some SNRs are dominated by IC or bremsstrahlung that contribute to flatter photon spectra than pions, the actual source proton index estimated from \textit{Fermi} SNR measurements would be steeper than the weighted average value of $s$. Particularly for the SNRs W44 and IC443 with clear characteristic pion-decay gamma-spectra, the observations give the accelerated proton index $s$ to be about 2.4 in the energy range smaller than break energy \citep{Ackermann et al. 2013}, where the projectile CRs in the galaxies mainly come from. Moreover, for TeV gamma-rays, we expect the signal is pionic and thus these index measurements can give us a fair estimate of the CR source index. The TeV data gives the index varies between 1.8-3.1 with an average value $s\sim 2.4$ \citep[e.g.,][]{Aliu et al. 2013,Brun et al. 2011,Aharonian et al. 2008}.

 For CR acceleration energy parameter, $\epsilon_{\rm cr}=E_{\rm sn}f_{\rm cr}$, the average kinetic energy released per SN ($E_{\rm sn}$) is $10^{51}{\rm erg}$ \citep{Woosley & Weaver 1995}, but there exists much uncertainties in the value of SNR acceleration efficiency to CR ($\epsilon _{\rm sn}$). If SNRs are the main sites of acceleration of cosmic rays, then 3 to 30$\ \rm{percent}$ of the supernova kinetic energy must end up transferred to CR protons from various theories: \citet{Fields et al. 2001} suggested that if SNRs are the dominant sources for cosmic-ray production as well as the nucleosynthesis of lithium, beryllium, and boron in the Milky Way, an acceleration efficiency of $\sim 30\ \rm{percent}$ is needed; \citet{Strong et al. 2010} obtains a CR energy input efficiency per SN of $3 - 10\ \rm{percent}$; \citet{Caprioli 2012}'s study also found the acceleration efficiency saturates at around $10-30\ \rm{percent}$; \citet{Dermer & Powale 2013}'s results suggest that most supernova remnants accelerate cosmic rays with an efficiency of $\sim 10\ \rm{percent}$ for the dissipation of kinetic energy into nonthermal cosmic rays. The observations of SNRs also give insight into CR acceleration efficiency, for example, SNR Tycho accelerates protons up to 500 TeV with an efficiency of $\sim 10\ \rm{percent}$ \citep{Morlino & Caprioli 2012} while the hadronic scenario of SNR RCW86 concludes that the accelerated particles energy efficiency from SNR is at the level of $\sim 0.07$ \citep{Lemoine-Goumard et al. 2012}.  We thus adopt a fiducial value $\epsilon_{\rm cr}=10\ \rm{percent}\times 10^{51}{\rm erg}=0.1\ \rm{foe}$, but note that uncertainties are large;
 we will adopt maximum value $\epsilon_{\rm cr,max}=0.3\ \rm{foe}$
 as implied by the Li, Be, and B nucleosynthesis results.

\section{Model Results}
\label{sec:Model Results}

\begin{figure}
\centering
\includegraphics[width=\columnwidth]{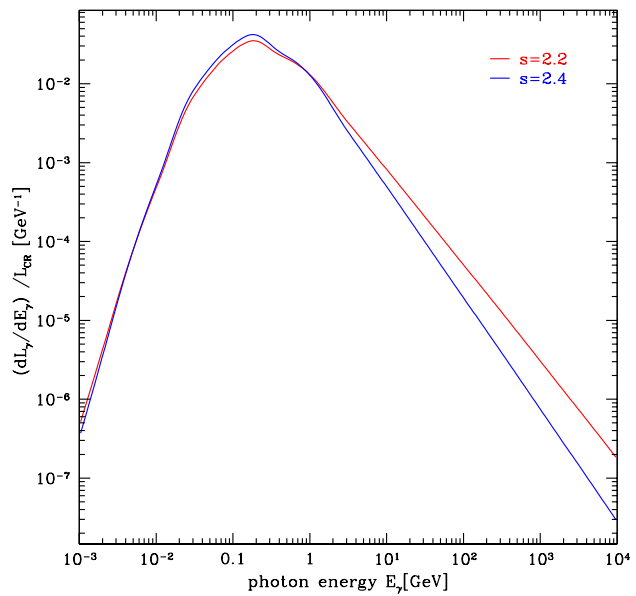}
\caption{\small
  Ratio of differential gamma-ray luminosity to total CR luminosity for a calorimetric galaxy. The red line represents the result with choice of source CR index $s=2.4$, while the blue line is for $s=2.2$.
\label{fig:spec-ratio}
}
\end{figure}

The thick-target model built in \S~\ref{sec:The Thick-Target/Calorimetric Model} gives proportionality relation of the differential gamma-ray emission to $\epsilon_{\rm cr}$, and from eq.~\ref{eq:gammalum}, we can see that $L_\gamma/L_{\rm CR}$ is the same for every calorimetric galaxy with the same choice of source CR proton index $s$, therefore 
\beq
\label{eq:dLoverL}
\frac{dL_\gamma/dE_\gamma}{L_{\rm CR}}
= \frac{{\cal A} E_\gamma dN_\gamma/dE_\gamma dt}{L_{\rm CR}}
\stackrel{\rm cal}{=} E_\gamma \frac{I_0(E_{\gamma},s)}{I(T_{\rm min})}{\rm GeV}^{-1}
 = const
\eeq
and the relation is shown in Fig.~\ref{fig:spec-ratio} with $s=2.2$ and 2.4. Because $(dL_\gamma/dE_\gamma)/L_{\rm CR}|_s$ is the same for all calorimetric galaxies, the plot of this ratio presents the general properties of our model's results: gamma-ray emission peaks around $\sim0.15{\rm GeV}$ and is nearly a power law at high energy. For different $s$, the ratios of differential gamma-ray luminosity to CR luminosity are different especially at high energy, but are always smaller than 1/3 due to energy conservation.

We now apply our model to individual starburst galaxies (\S \ref{subsec:individual starbursts}).
With their cosmic-ray parameters determined, we then compute their luminosity and
evaluate their status as calorimeters (\S \ref{subsec:Calorimetric Limit}).

\subsection{Individual Starbursts}
\label{subsec:individual starbursts}

We now apply our  model to five individual starbursts NGC 253, M82, NGC 4945, NGC 1068, and the Circinus galaxy, as well as the ULIRG Arp 220.
The input parameters and best-fit results are listed in Table~\ref{tab:parameter} and Table~\ref{tab:result}. For each galaxy we adopt an observed star-formation rate (SFR), and then calculate the pionic flux $E_{\gamma}^2dN_{\pi\to\gamma\gamma}/dE_{\gamma}dAdt$ for each point in $(\epsilon_{\rm cr},s)$ space. We perform $\chi^2$ test with the observed gamma-ray data to get the best-fit model parameters:
\beqar
\label{eq:chi-squared}
\chi^2(\epsilon_{\rm cr},s)=\sum\limits_{i}\frac{(F_{\rm i}-\hat{F_{\rm i}})^2}{\sigma_{\rm i}^2}
\eeqar
where $\hat{F_{\rm i}}$ is the flux value of the data points at each photon energy $E_{\rm i}$, $F_{\rm i}=\epsilon_{\rm cr}y_{\rm i}(s)$ is our model's flux value at each $E_{\rm i}$, $\sigma_{\rm i}$ is the uncertainty of the data's flux value at each $E_{\rm i}$. 

\begin{table*}
\caption{Parameters set for the starburst galaxies in Thick-Target Model. 
\label{tab:parameter}
}
\begin{tabular}{c|ccc|cc}
\hline\hline
Galaxy & Distance
  & SFR $\psi$ & SN Rate $R_{\rm SN}$ & GeV data & TeV data\\
Name & $D \rm [Mpc]$ & $[M_\odot/\rm yr]$ & $[\rm century^{-1}]$ & reference & reference\\
\hline
M82 & $3.4\pm0.9$ & $6.3\pm0.9$ & $5.7\pm0.9$ & \citet{Acker2012} & \citet{VERITAS Collaboration et al. 2009} \\
NGC 253 & $2.5\pm0.5$ & $2.9\pm0.4$ & $2.6\pm0.4$ & \citet{PA2012} & \citet{Abramowski et al. 2012}\\
NGC 4945 & $3.7\pm0.8$ & $3.5\pm1.0$ & $3.2\pm0.9$ & \citet{Acker2012} &  \\
NGC 1068 & $16.7\pm3.0$ & $38\pm10$ & $35\pm9$ & \citet{Acker2012} & \citet{Ah2005} \\
Circinus & $4.2\pm0.7$ & $2.1\pm0.5$ & $1.9\pm0.5$ & \citet{Hayashida et al. 2013} &  \\
Arp 220 & $77.0\pm2.0$ & $188.3\pm10.0$ & $172.1\pm9.1$ & \citet{Peng 2016} & \citet{Fleischhack 2015}\\
\hline\hline
\end{tabular}\\
{
Galaxy distances: \citet{GS2004}. Star formation rates: \citet{Acker2012}, using \citet{GS2004} total IR luminosities and Kennicutt relation \citep{Kennicutt 1998}, except for the Circinus galaxy \citep{Tully2009, Hayashida et al. 2013} and Arp 220 \citep{Peng 2016}. \\
Distance uncertainties come from
  \url{http://ned.ipac.caltech.edu}. Except for the ULIRG Arp 220, the redshift-dependent distance uncertainty comes from Hubble constant error \citep{Planck 2016} (assuming the peculiar velocity uncertainty is the same as Hubble constant uncertainty).}
\end{table*}

\begin{table*}
      \caption{ Results for the starburst galaxies in Thick-Target Model.
\label{tab:result}
}
\begin{tabular}{c|ccc|cc}
\hline\hline
Galaxy & CR source  & CR acceleration energy & $L_{\rm 0.1-100GeV}^{model}$ & $L_{\rm 0.1-100GeV}^{\rm Fermi}$ & $\Gamma^{\rm Fermi}$ \\
Name &  index $\hat{s}$ & per SN $\epsilon_{\rm cr} [{\rm foe/SN}] $ & $\rm [10^{40}erg s^{-1}]$ & $\rm [10^{40} erg s^{-1}]$ &  \\
\hline
M82 & $2.275\pm0.102$ & $0.106\pm0.025$ & $1.48\pm0.44$ & $1.47\pm0.14$ & $2.25\pm0.13$\\
NGC 253 & $2.350\pm0.037$ & $0.116\pm0.013$ & $0.73\pm0.10$ & $0.60\pm0.07$ & $2.18\pm0.09$\\
NGC 4945 & $2.400\pm0.446$ & $0.210(>0.103)$ & $1.64(>0.80)$ & $1.17\pm0.23$ & $2.05\pm0.13$ \\
NGC 1068 & $2.100\pm0.617$ & $0.253(>0.128)$ & $13.3(>6.7)$ & $15.0\pm2.9$ & $2.29\pm0.19$\\
Circinus & $2.300\pm0.486$ & $0.619(>0.310)$ & $2.97(>1.48)$ & $2.9\pm0.5$ & $2.19\pm0.12$ \\
Arp 220 & $2.550\pm0.257$ & $0.808(>0.404)$ & $2.85(>1.43) \times 10^2$ & $(1.78\pm0.3)\times 10^2$ & $2.35\pm0.16$ \\
\hline\hline
\end{tabular}\\
{
  {\em Fermi} gamma luminosities for the galaxies are calculated by \citet{Hayashida et al. 2013} using a power law spectral model $dN/dE \propto E^{-\Gamma}$, except for Arp 220 \citep{Peng 2016}.
  }
\end{table*}

We consider injection indices in the range $s \in [2.1,3.0]$.
By maximizing the value of $\chi^2$ at each $s$, we can get the best-fit values of $\epsilon_{\rm cr}$ analytically.  We then compare the values of $\chi^2$ for each $s$ with the best-fit $\epsilon_{\rm cr}$, finally can find the best-fit value of $s$ numerically.

From Table~\ref{tab:result}, we can see that the pionic gamma-ray luminosity calculated from our model agrees well with the phenomenological \textit{Fermi} fits for the starburst galaxies M82, NGC 253, NGC 4945, NGC 1068 the Circinus galaxy, and ULIRG Arp 220.

\begin{figure*}
\centering
\includegraphics[width=\columnwidth]{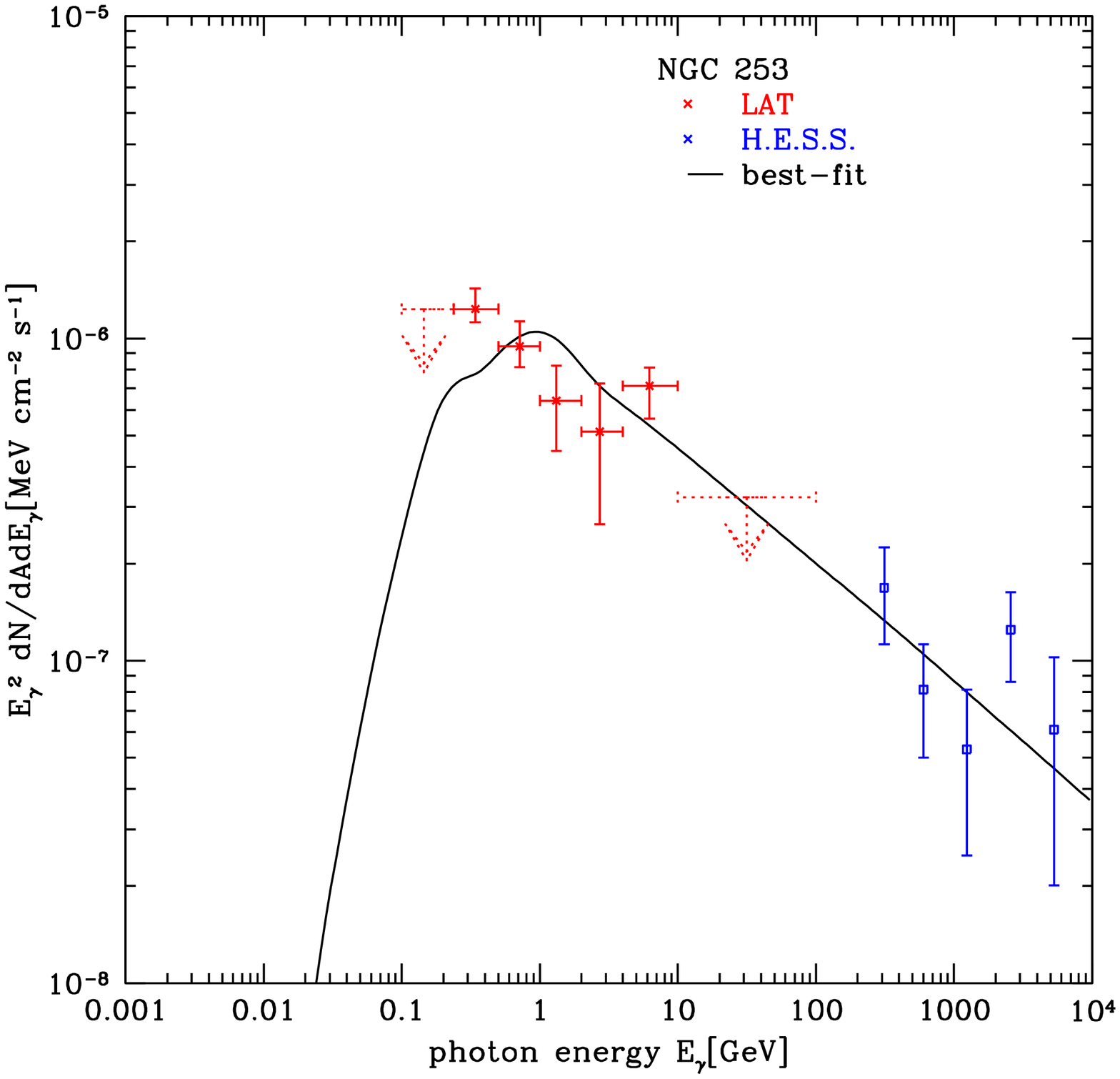}
\includegraphics[width=\columnwidth]{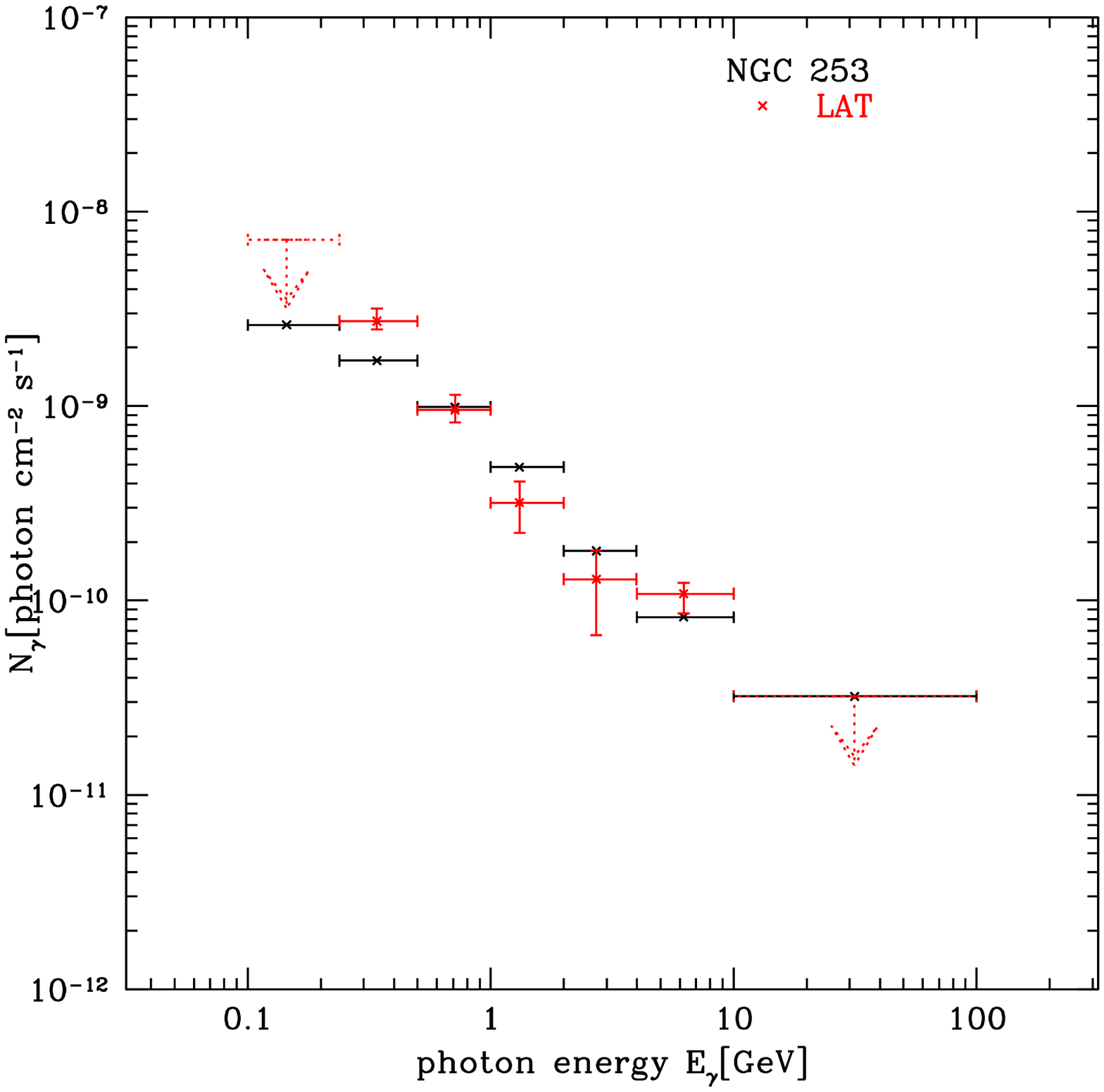}
\caption{\small
Left panel: Differential pionic gamma-ray spectrum (solid curve) for NGC 253 with the best-fit parameters: source CR index $s$ and accelerated CR energy per SN $\epsilon_{\rm cr}$. {\em Fermi} points are stars (red), {\em H.E.S.S} points are squares (blue), black solid line is our model's best-fit to data; see Table~\ref{tab:result}. Right panel: Best-fit integrated pionic gamma-ray spectrum for NGC 253. Red points are {\em Fermi} measurement, black points are our model's best-fit results.
\label{fig:spec-NGC253}
}
\end{figure*}

\begin{figure*}
\centering
\includegraphics[width=\columnwidth]{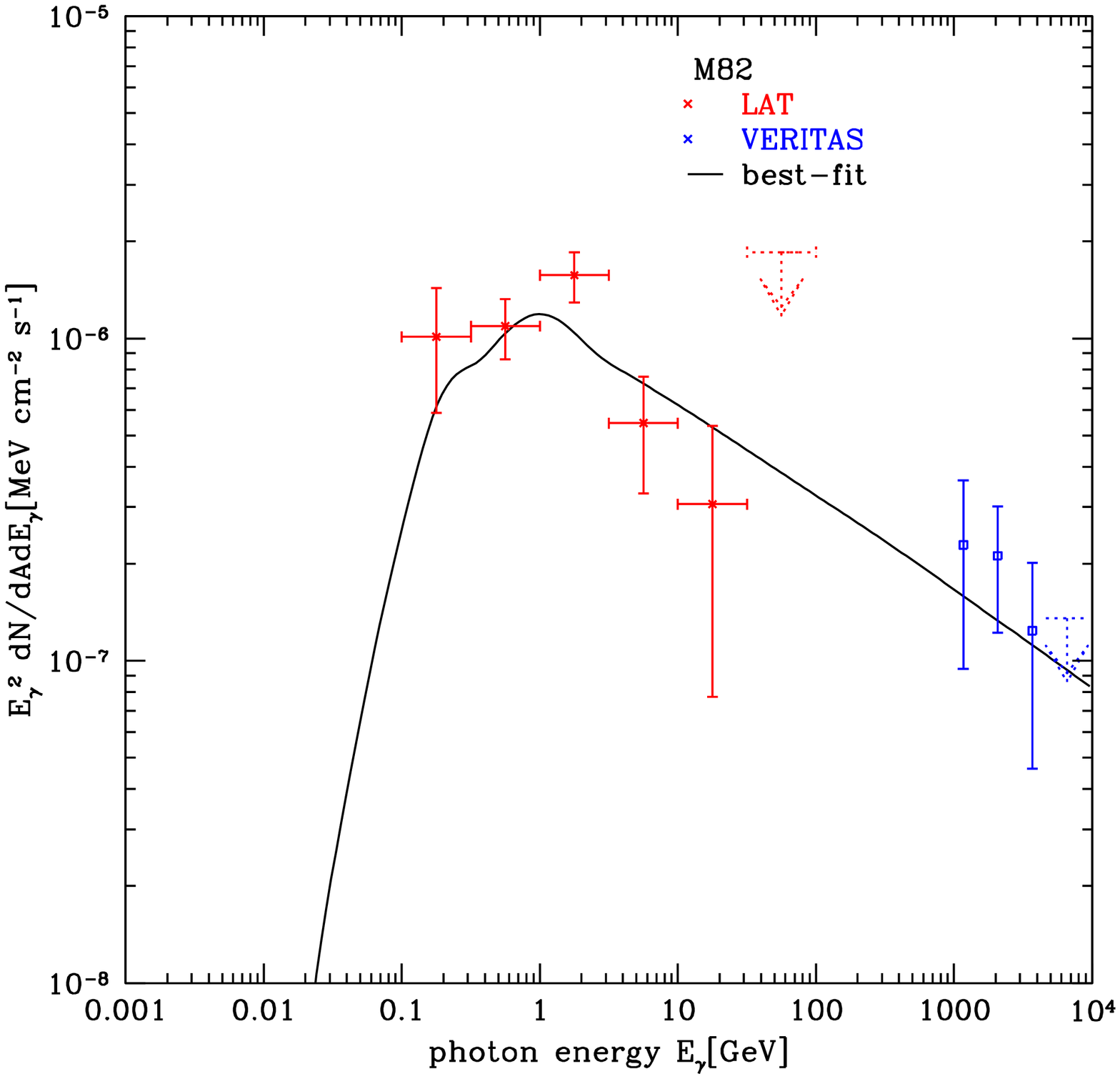}
\includegraphics[width=\columnwidth]{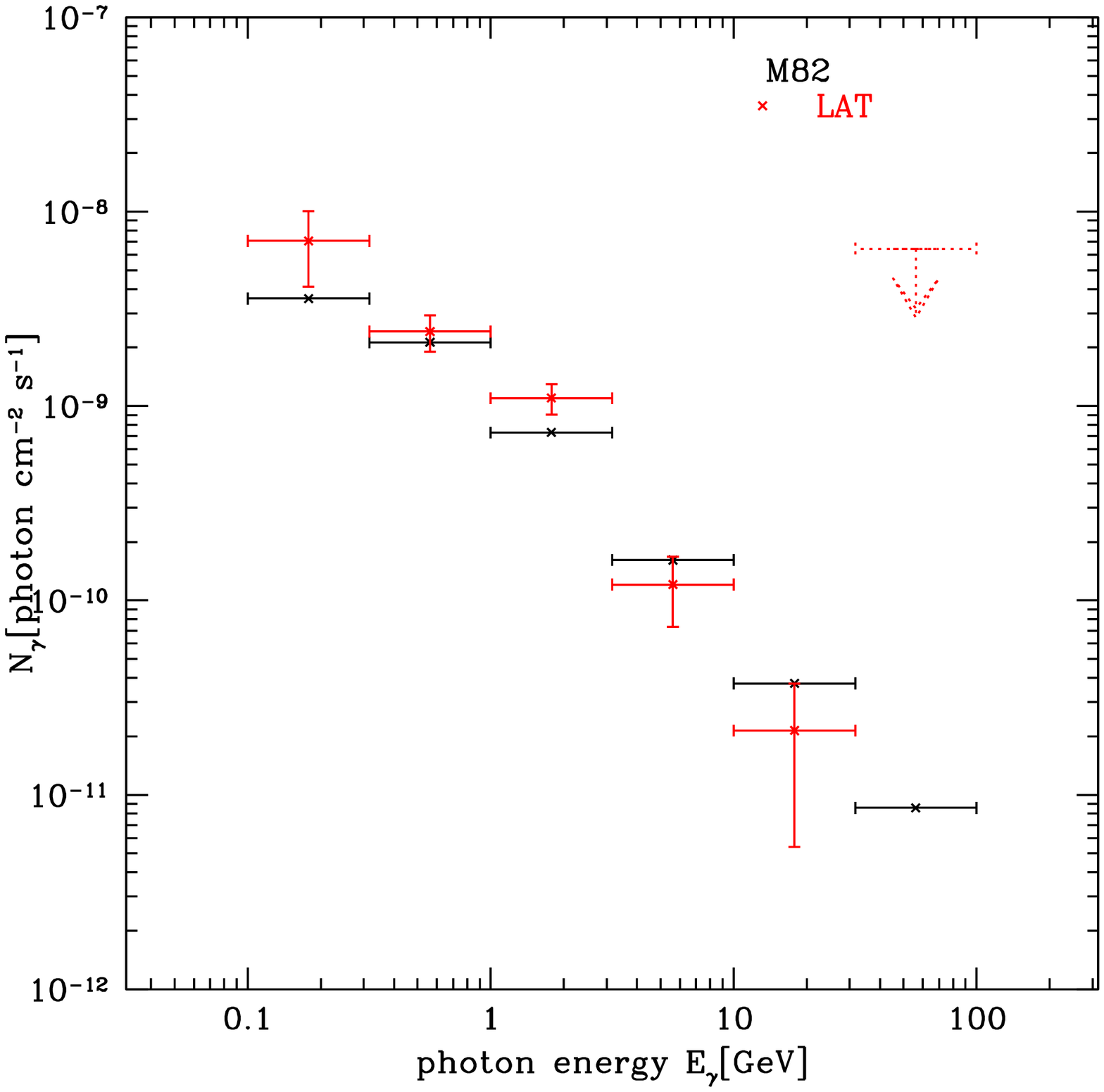}
\caption{\small
Left panel: Differential pionic gamma-ray spectrum (solid curve) for M82 with the best-fit parameters: source CR index $s$ and accelerated CR energy per SN$\epsilon_{\rm cr}$. {\em Fermi} points are stars (red), {\em VERITAS} points are squares (blue), black solid line is our model's best-fit to data; see Table~\ref{tab:result}. Right panel: Best-fit integrated pionic gamma-ray spectrum for M82. Red points are {\em Fermi} measurement, black points are our model's best-fit results.
\label{fig:spec-M82}
}
\end{figure*}

\begin{figure*}
\centering
\includegraphics[width=\columnwidth]{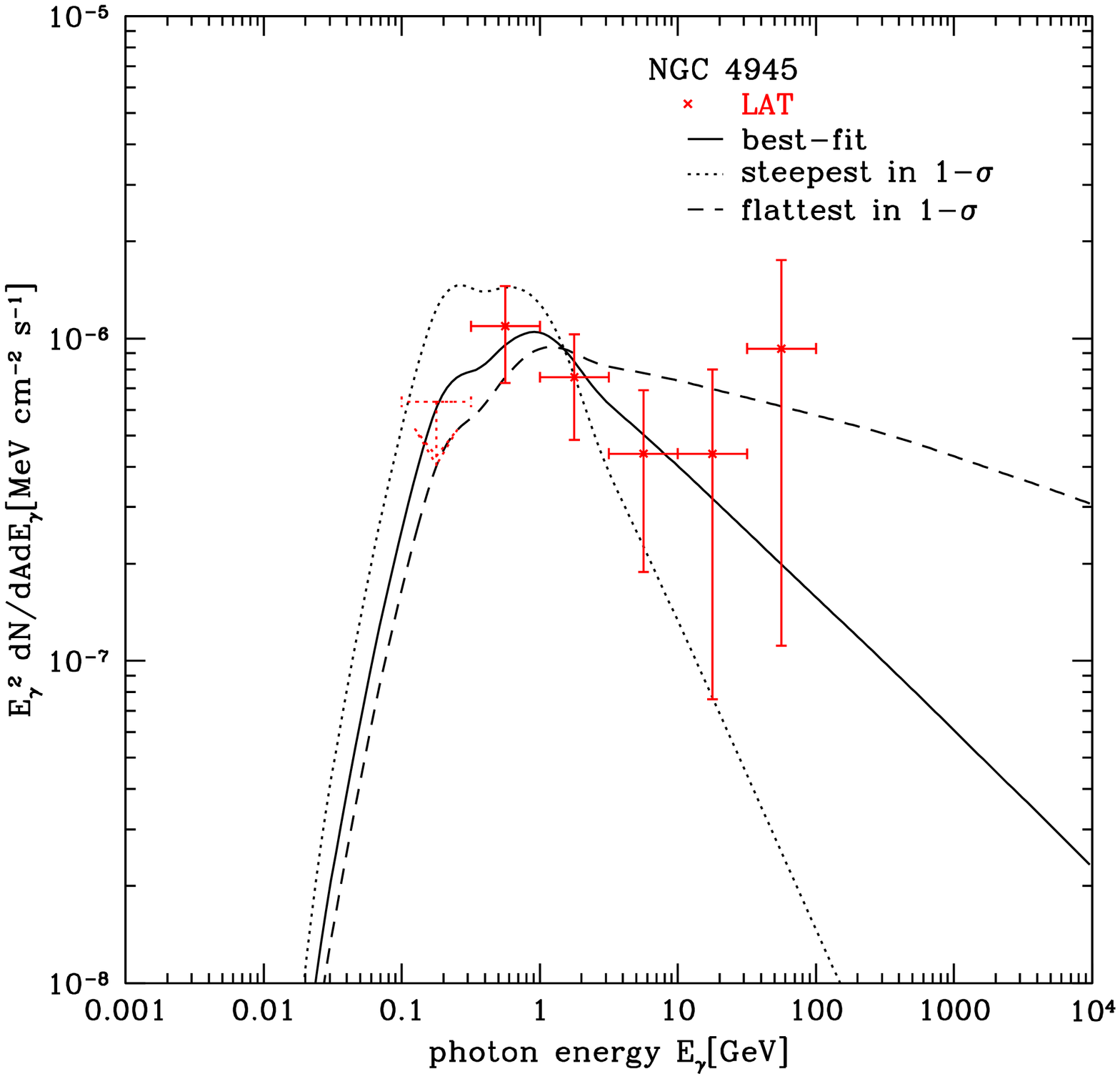}
\includegraphics[width=\columnwidth]{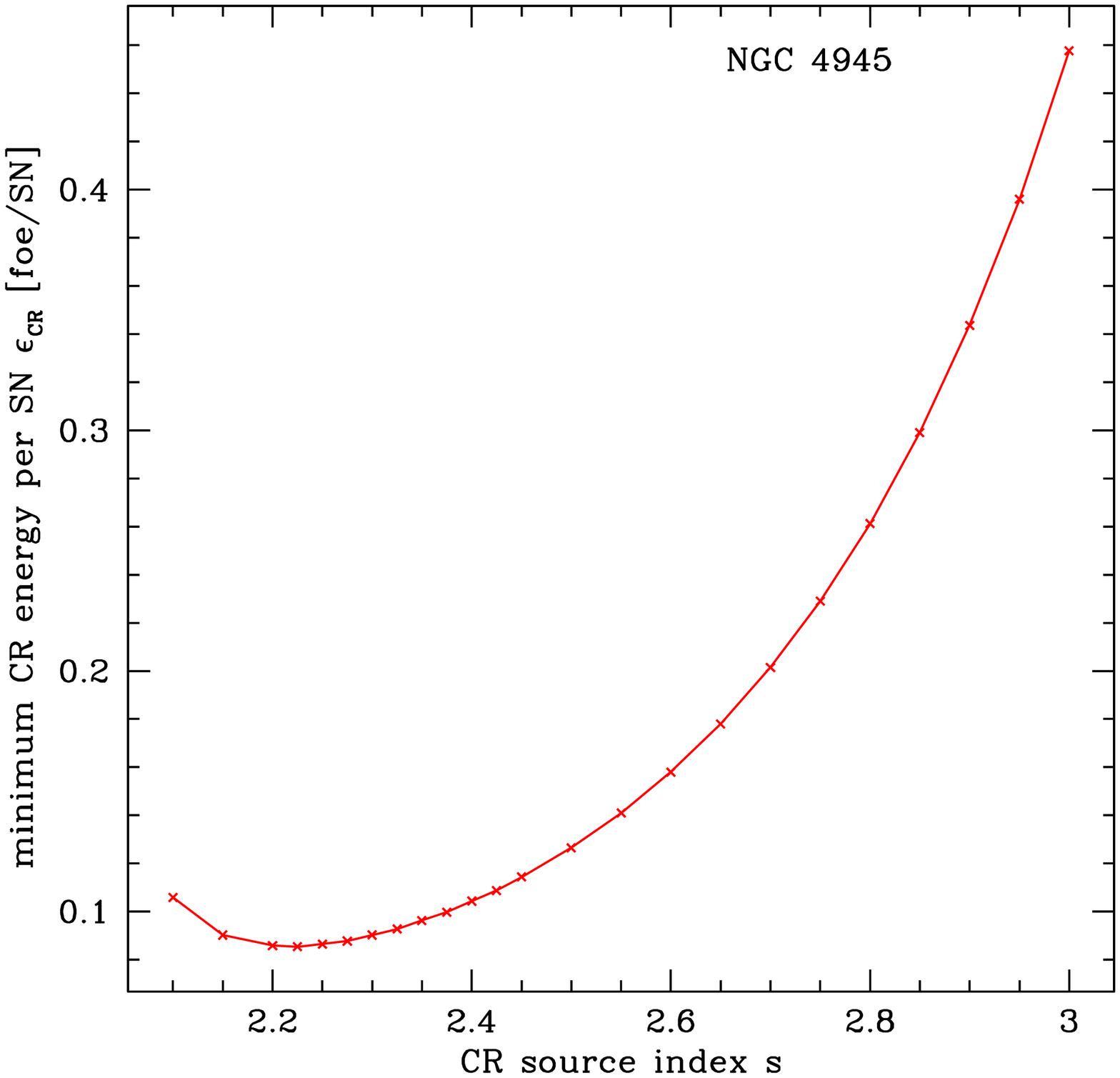}\\
\includegraphics[width=\columnwidth]{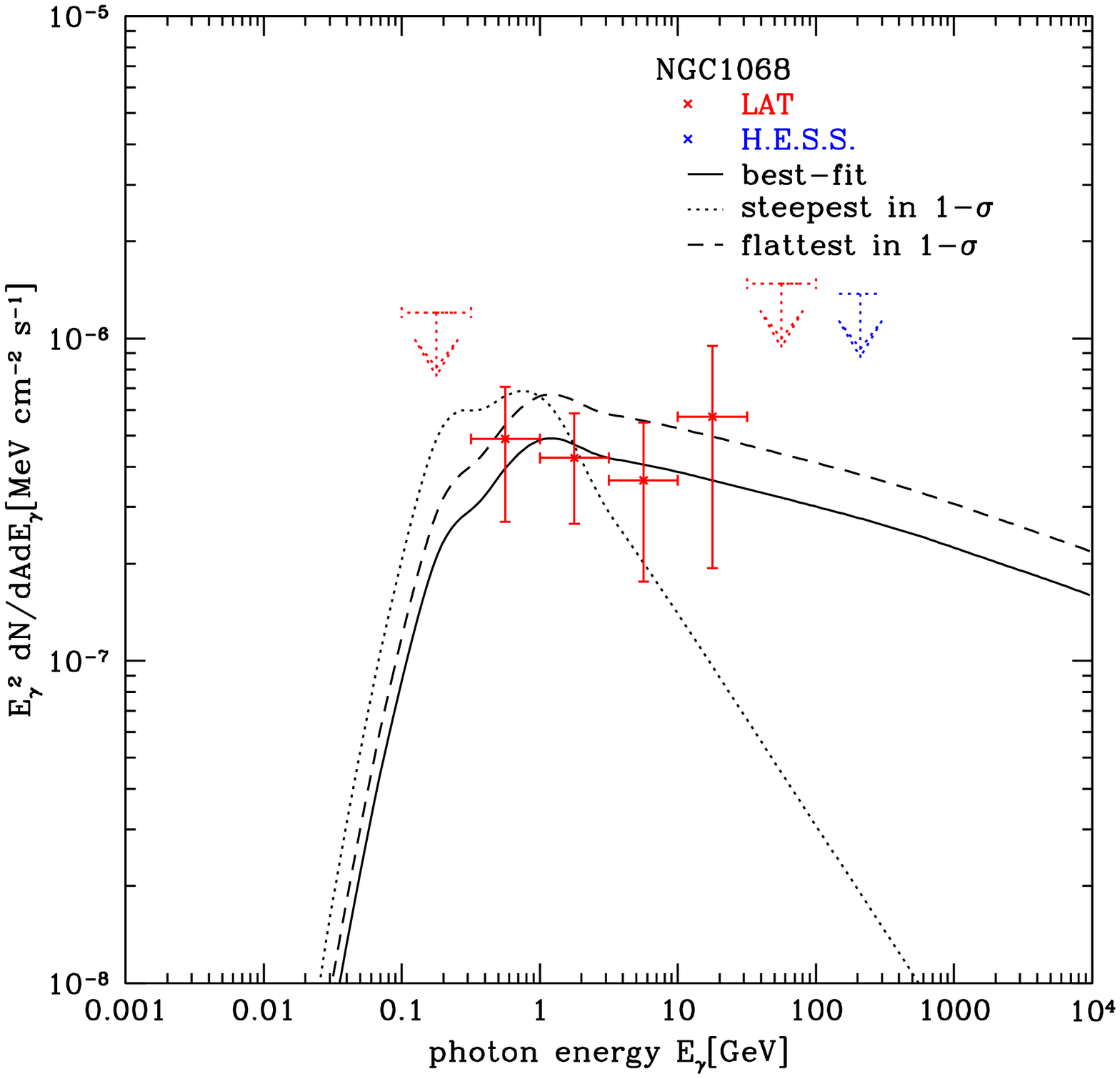}
\includegraphics[width=\columnwidth]{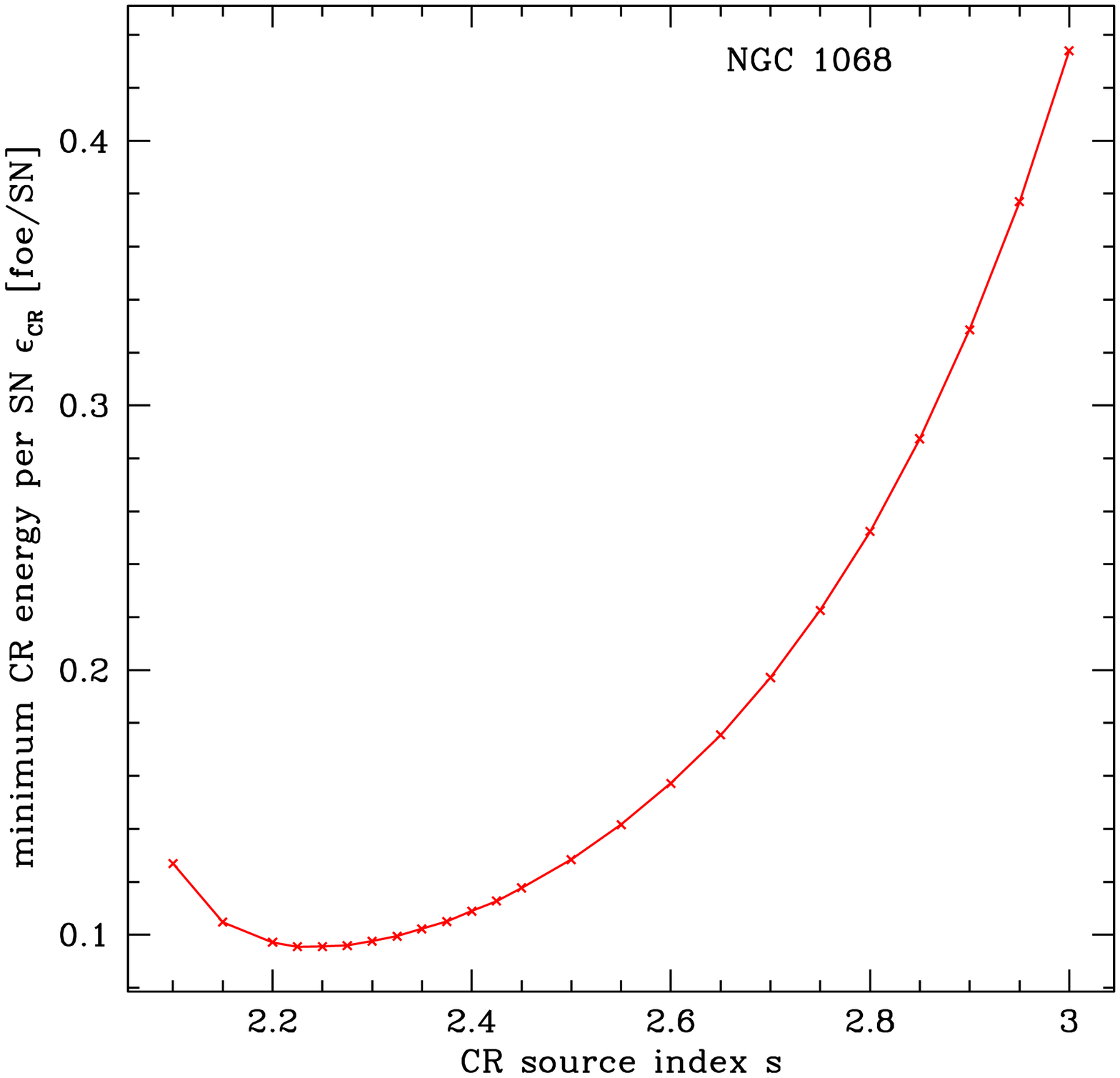}\\
\caption{\small
Left panel: Pionic gamma-ray spectra (solid curve) for NGC 4945 (upper) and NGC 1068 (lower) with the best-fit parameters: source CR index $s$ and accelerated CR energy per SN $\epsilon_{\rm cr}$. {\em Fermi} points are stars (red), {\em H.E.S.S.} points are squares (blue), black solid line is our model's best-fit to data; see Table~\ref{tab:result}.  Black dashed line is our model's flattest curve to fit the data in 1-$\sigma$ error, while black dotted line is the steepest curve in 1-$\sigma$ error, the parameters' values of these curves are the cross points in Fig.~\ref{fig:contour}.
Right panel: minimum $\epsilon_{\rm cr}$ vs. $s$ for NGC 4945 (upper), NGC 1068 (lower).
\label{fig:spec-stb}
}
\end{figure*}

\begin{figure*}
\centering
\includegraphics[width=\columnwidth]{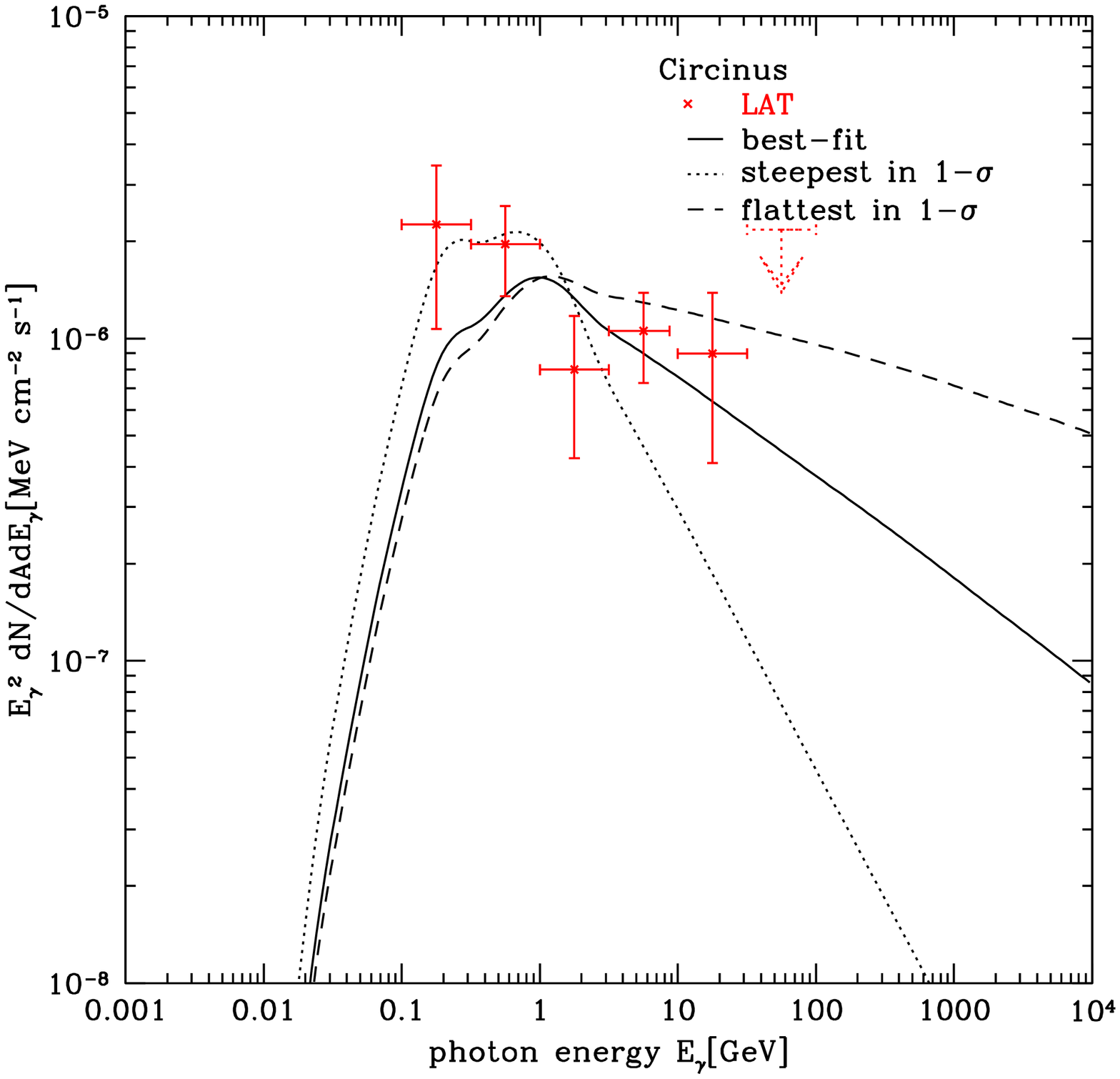}
\includegraphics[width=\columnwidth]{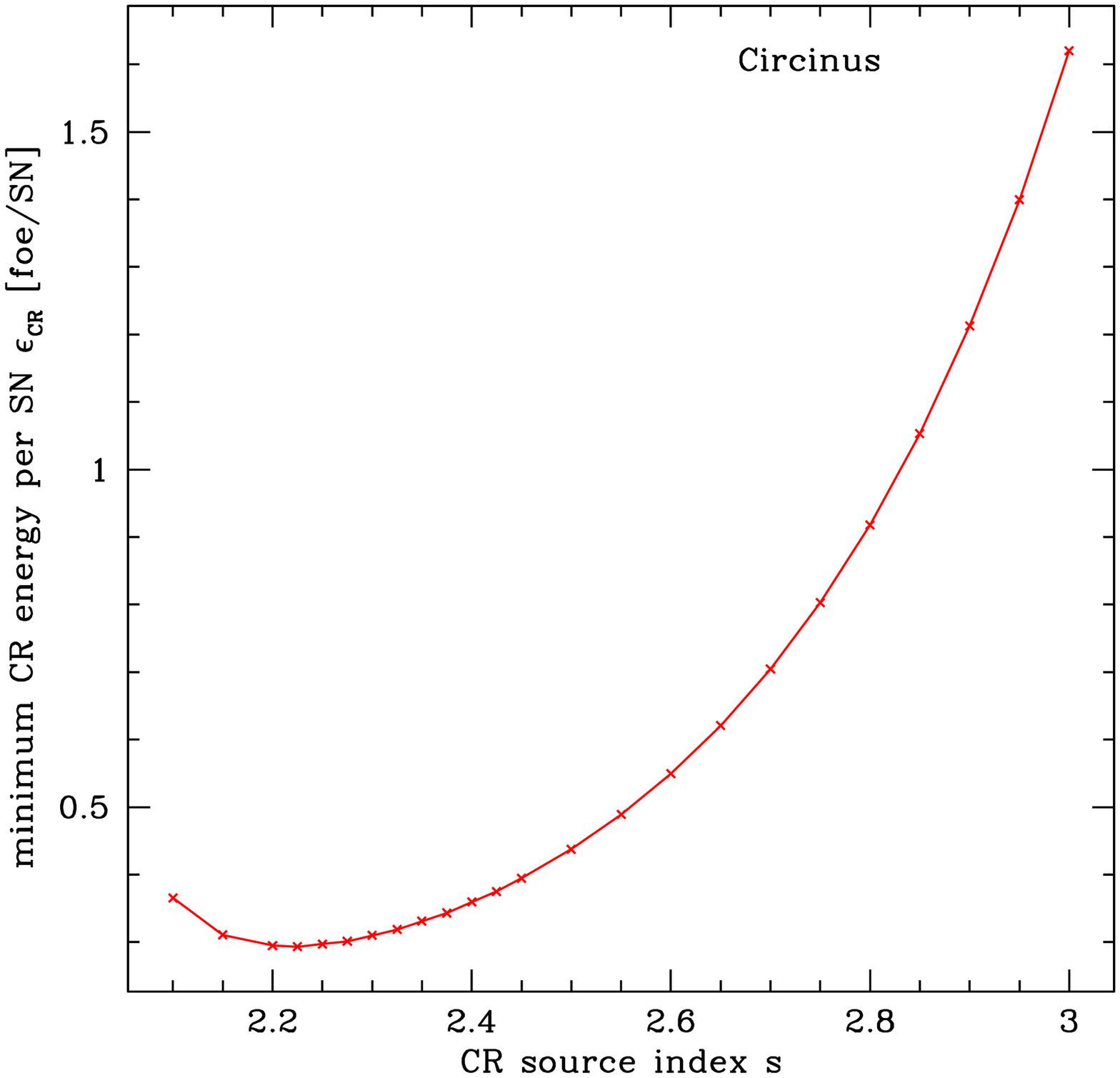}\\
\includegraphics[width=\columnwidth]{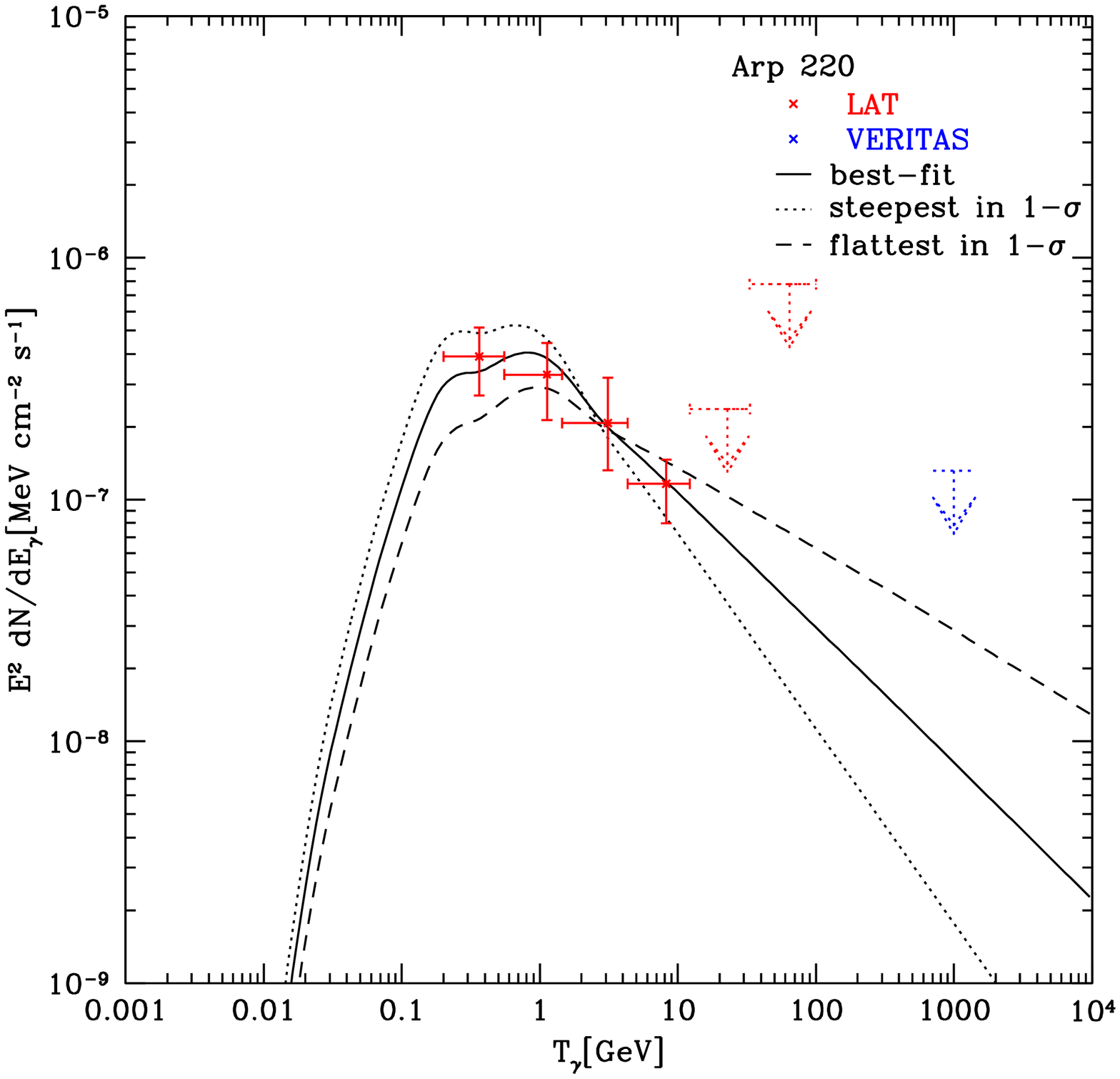}
\includegraphics[width=\columnwidth]{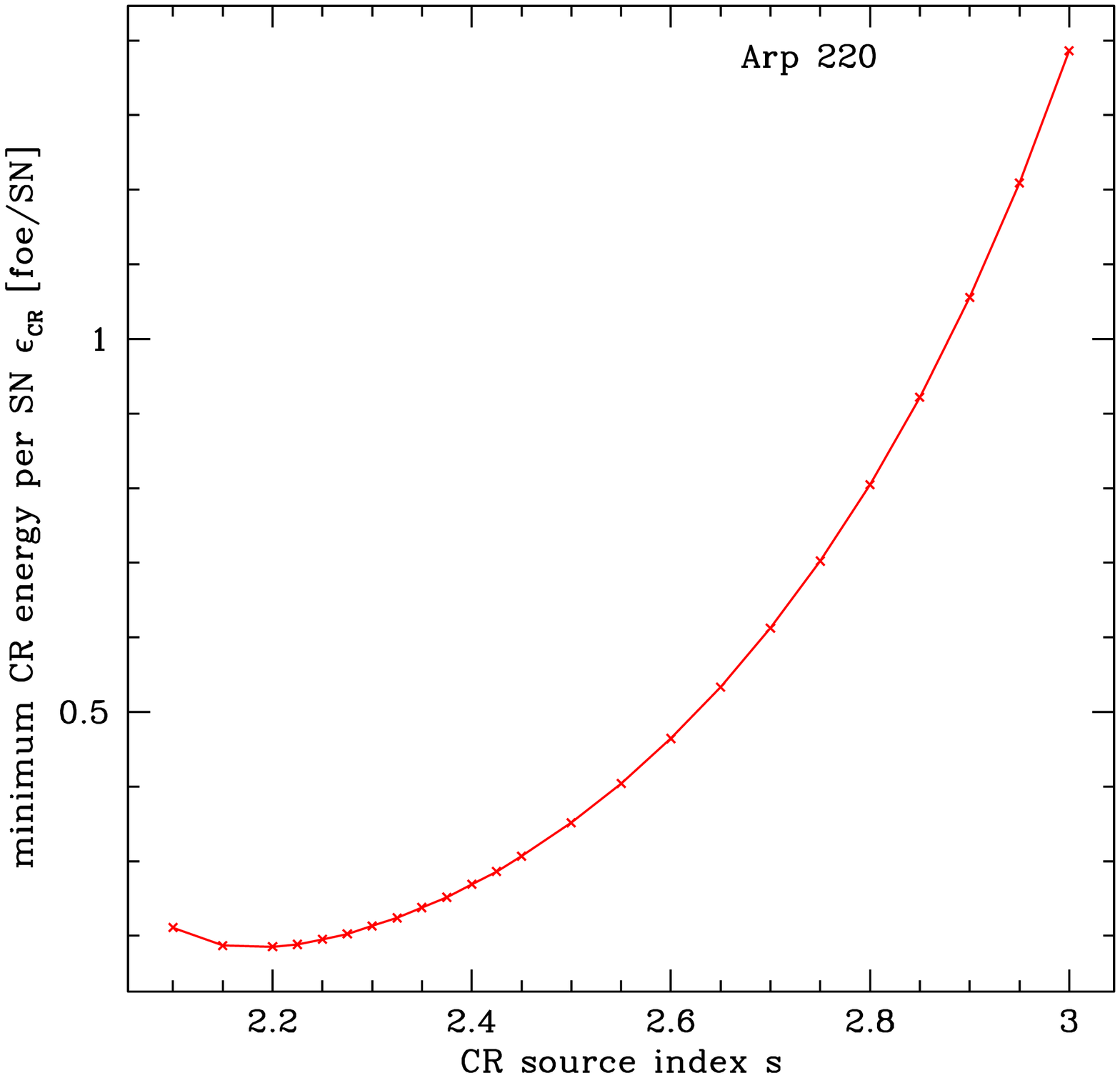}
\caption{\small
Left panel: Pionic gamma-ray spectra (solid curve) for the Circinus galaxy (upper) and Arp 220 (lower) with the best-fit parameters: source CR index $s$ and accelerated CR energy per SN $\epsilon_{\rm cr}$. {\em Fermi} points are stars (red), {\em VERITAS} points are squares (blue), black solid line is our model's best-fit to data; see Table~\ref{tab:result}. Black dashed line is our model's flattest curve to fit the data in 1-$\sigma$ error, while black dotted line is the steepest curve in 1-$\sigma$ error, the parameters' values of these curves are the cross points in Fig.~\ref{fig:contour}.
Right panel: minimum $\epsilon_{\rm cr}$ vs. $s$ for the Circinus galaxy (upper) and Arp 220 (lower).
\label{fig:spec-stb_2}
}
\end{figure*}

The best-fit pionic gamma-ray spectra can be seen in Figs.~\ref{fig:spec-NGC253} through \ref{fig:spec-stb_2}. In left panels, the solid lines is our model's calculated differential spectral energy distribution of the five starburst galaxies with the best-fit parameters $s$ and $\epsilon_{\rm cr}$. The red points in GeV range are \textit{Fermi} data while blue ones in TeV range are got from \textit{H.E.S.S} or \textit{VERITAS}. For M82 and NGC 253, we see that our best fit to GeV and TeV data is quite good and fairly well constrained thanks to the relatively large energy range. For NGC 1068, NGC 4945, Circinus and Arp 220, only GeV data is available and even our simple model is poorly constrained.

We note that the observed differential spectrum points are derived
assuming a constant spectral index at all energies,
but in our model the index varies strongly at lower energies
near the ``pion bump'' at $m_{\pi^0}/2$.
We thus plot in the
right panels Figs.~\ref{fig:spec-NGC253}--\ref{fig:spec-stb_2}
the integrated photon flux $\int_{E_i^{\rm min}}^{E_i^{\rm max}} dF/dE \ dE$ 
over each energy bin $i$,
whose width is spanned by the horizontal bars.
This corresponds to the photon counts per energy
bin, which is what {\em Fermi} directly measures
and which is free from assumptions about spectral index.
The black points are from our best-fit model, and the red points are the
{\em Fermi} data.
We see that our fits are generally
good across the GeV range, including
at low energies near the pion bump where the spectral index is not constant.

From Figs.~\ref{fig:spec-NGC253}--\ref{fig:spec-stb_2}, we can see that the gamma-ray spectra got from our thick-target model has the following features, as already seen in Fig.~\ref{fig:spec-ratio}: (1) the shape only depends on the injected proton spectrum; (2) the magnitude is proportional to $\epsilon_{\rm cr}$; (3) at high energies, the gamma-ray spectral index is the same as the proton injection index $s$; (4) in our model, the peak is due to the pion bump,
  which appears at $E_{\gamma} = m_{\pi^0} = 67.5 \ \rm MeV$ in plots of $F_E$,  \citep{Stecker 1971, Dermer 1986}, but is shifted to $\sim 1$ GeV in our
  $E^2 F_E$ plots.

\begin{figure*}
\centering
\includegraphics[width=0.7\columnwidth]{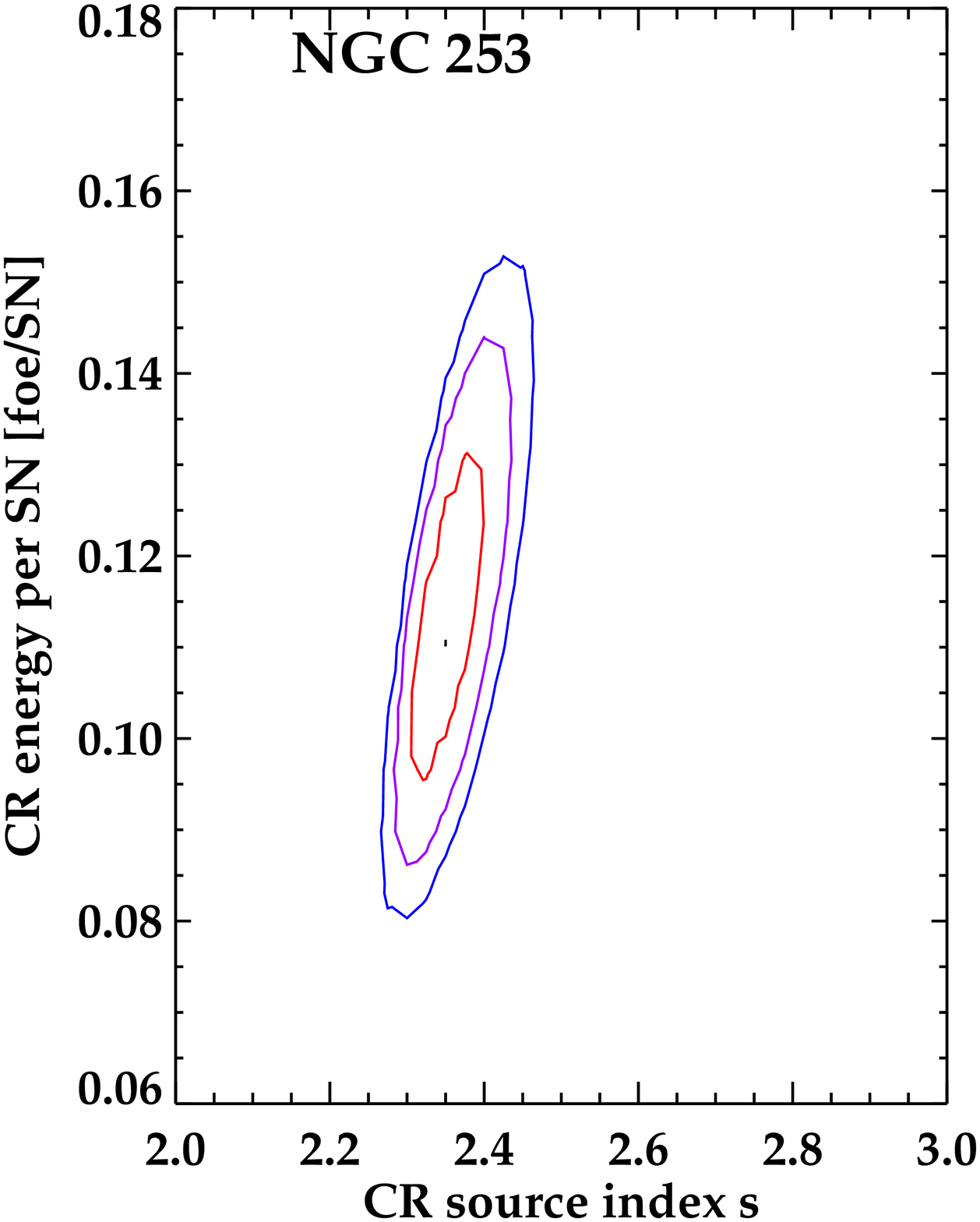}
\includegraphics[width=0.7\columnwidth]{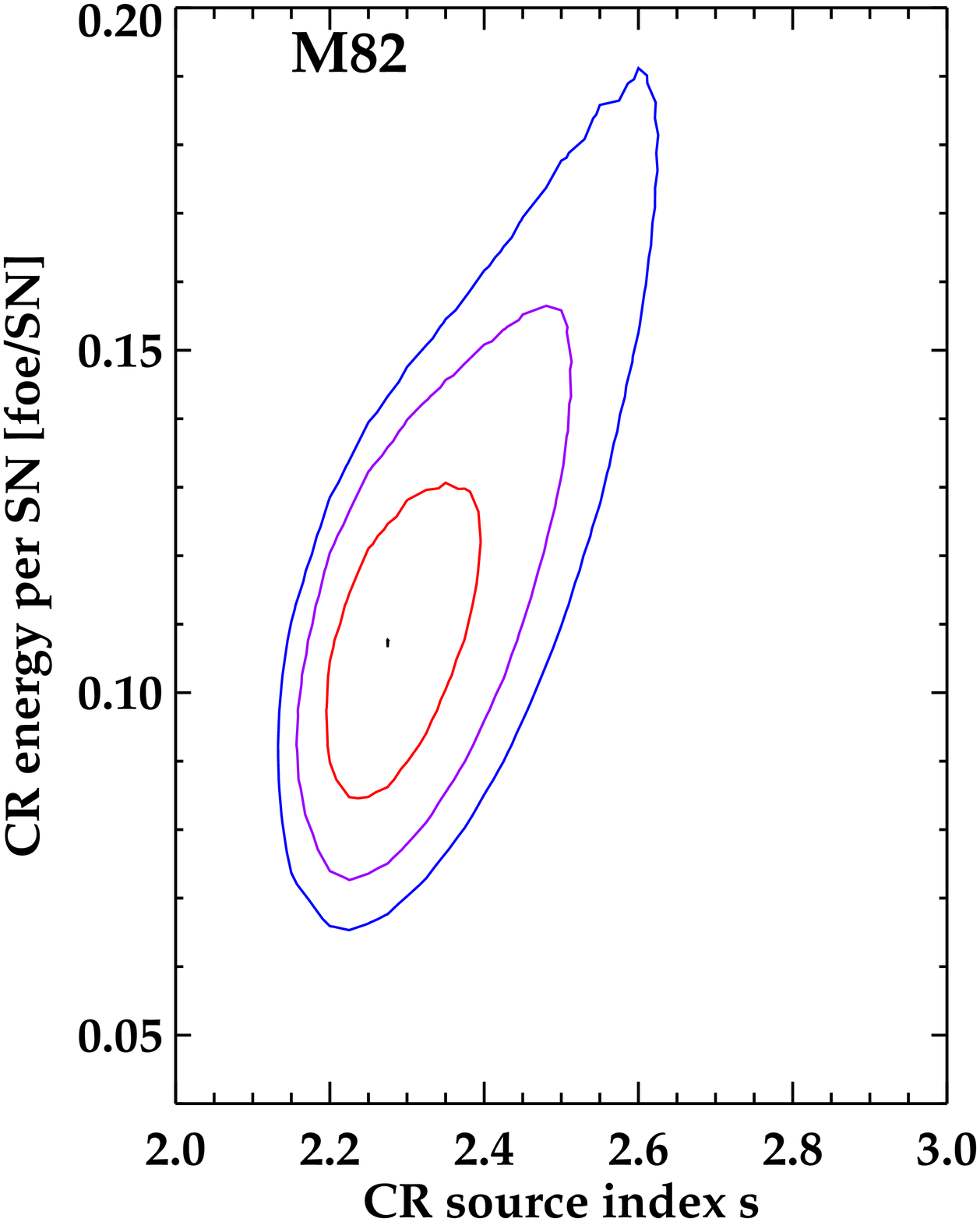}\\
\includegraphics[width=0.7\columnwidth]{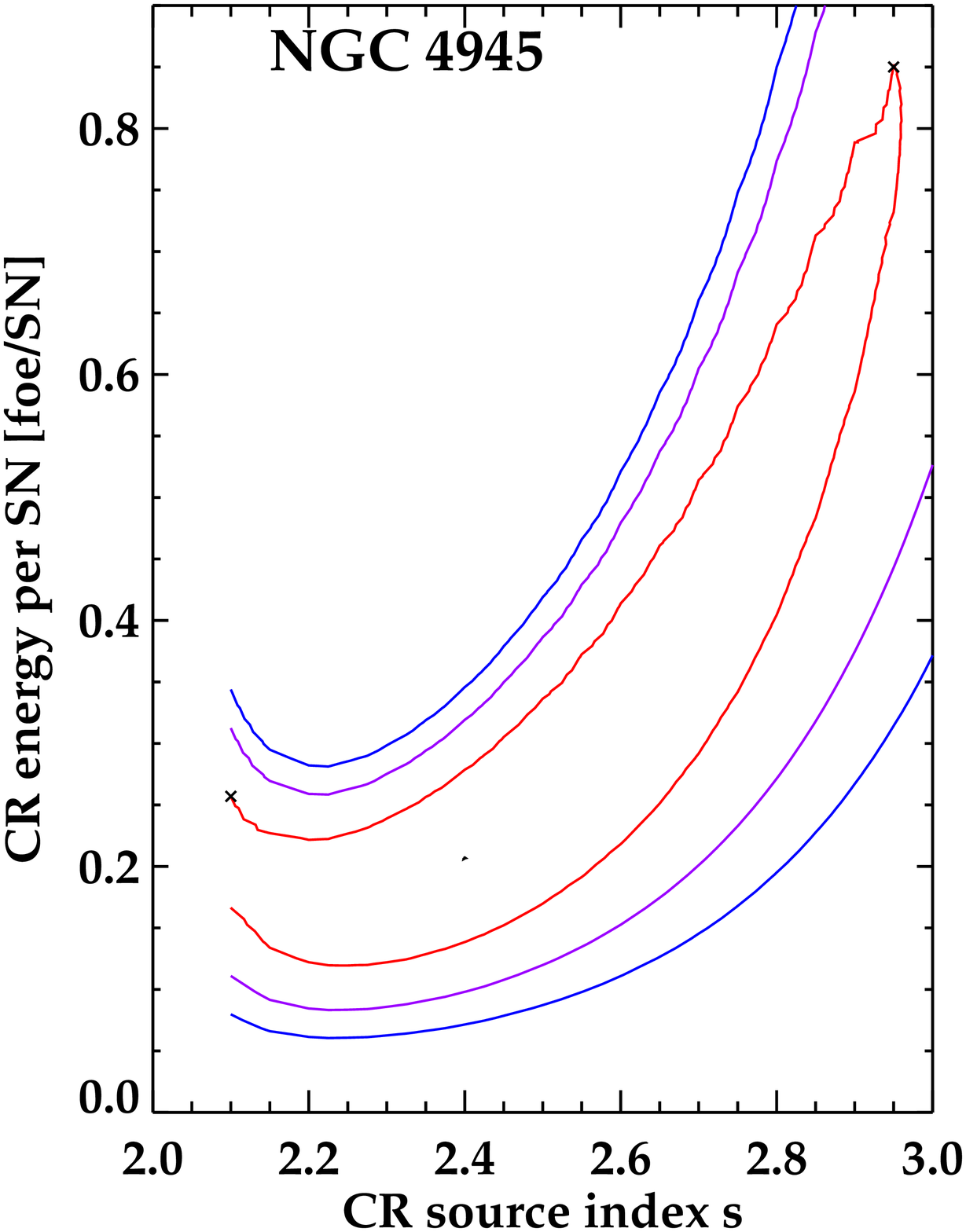}
\includegraphics[width=0.7\columnwidth]{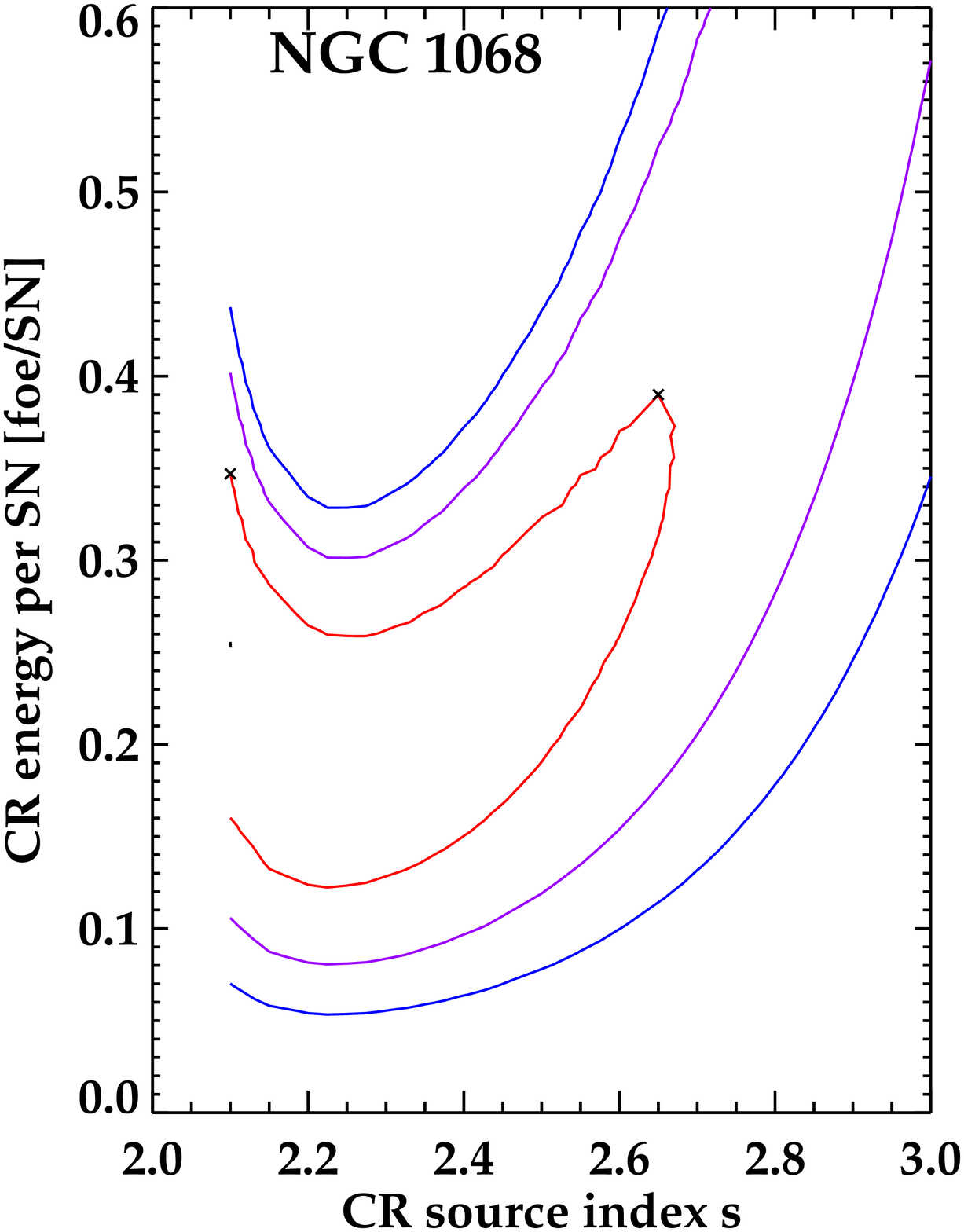}\\
\includegraphics[width=0.7\columnwidth]{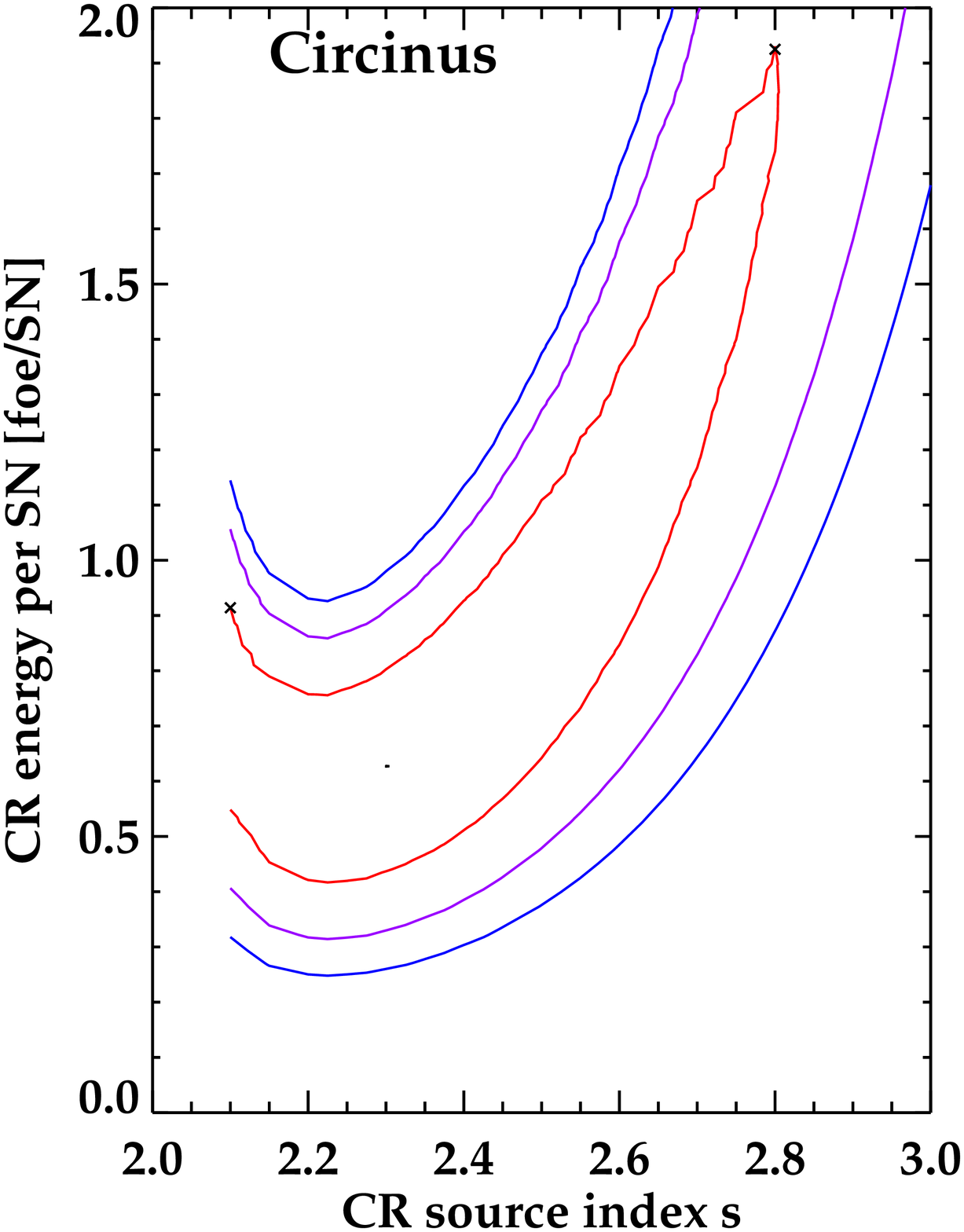}
\includegraphics[width=0.7\columnwidth]{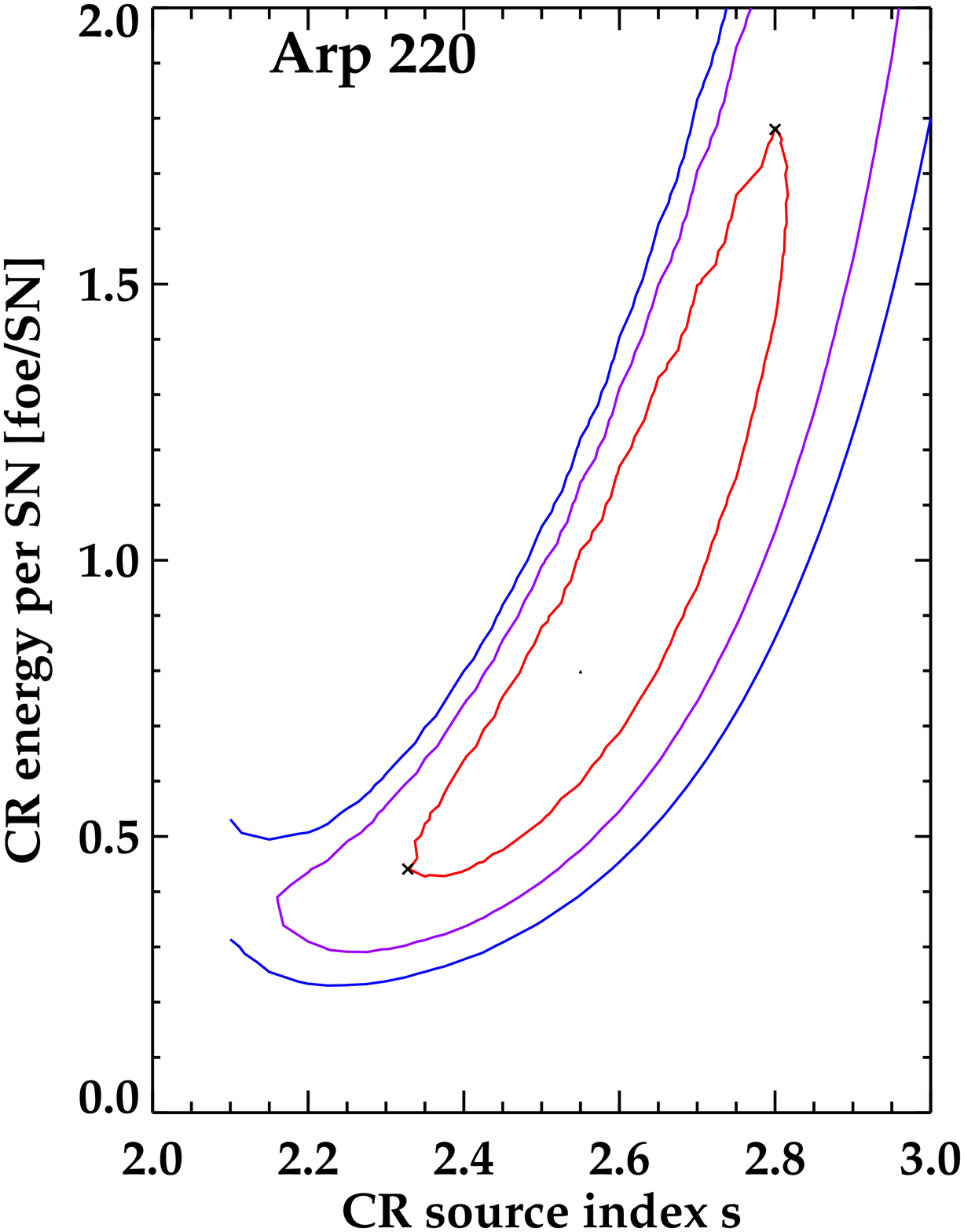}
\caption{\small
  Contour plots of $\chi^2$ for our model fits to starburst galaxy data.
  The best-fit values are the central black dot; (red, magenta, blue) lines represent ($70\ \rm{percent}$ CL, $95\ \rm{percent}$ CL, $99\ \rm{percent}$ CL). For starbursts without TeV data and the ULIRG Arp 220, the 1-$\sigma$ fit values of the flattest and steepest curves are the cross points, the corresponding curves are shown in Fig.~\ref{fig:spec-stb} and Fig.~\ref{fig:spec-stb_2}. 
\label{fig:contour}
}
\end{figure*}

The $\chi^2$ contour plots are shown in Fig.~\ref{fig:contour} with Confidence Level (CL) $=(70\ \rm{percent},95\ \rm{percent},99\ \rm{percent})$.  For M82 and NGC 253, TeV data and good GeV data are available,
and $s$ and $\epsilon_{\rm cr}$ are both well-constrained. For these galaxies,
$\epsilon_{\rm cr}\sim 0.1 \ \rm foe$, in good agreement with canonical estimates for Milky-Way cosmic rays \citep{Ginzburg & Syrovatskii 1964}. We see
that steeper (shallower) $s$
can be accommodated by a higher (lower) $\epsilon_{\rm cr}$.  This
arises physically because
$\epsilon_{\rm cr}$ fixes the overall normalization, and 
thus to fit 
the high-energy data with a steeper slope requires a higher overall
normalization.  The resulting tension with the low-energy points
limits the range of this correlation.

For the other galaxies NGC 1068, NGC 4945, Circinus and Arp 220,
the lack of TeV data leaves
large uncertainties in {\em both} $s$ and $\epsilon_{\rm cr}$,
as seen in Fig.~\ref{fig:contour}.
But Fig.~\ref{fig:contour} nevertheless shows that
GeV data place a lower bound on $\ecrsn$. 
Using the $\chi^2$ to find the likelihood function $P$, we compute
$P(>\epsilon_{\rm cr,min}|s)=95\ \rm{percent}$ to derive the
$95\ \rm{percent}$ CL lower limit $\epsilon_{\rm cr,min}$ to the supernova energy
per supernova for each value of acceleration index $s$.
Results appear in the left panels of Figs.~\ref{fig:spec-stb} and \ref{fig:spec-stb_2}, where we see that $\epsilon_{\rm cr,min}$
is always at its smallest values for $s \sim 2.2$, i.e., the
preferred theoretical and Milky-Way value.  But 
as $s$ increases, $\ecrsn$ becomes quite large.
This reiterates that TeV data for these starburst
is critical to CR spectral index $s$ and thus getting better-constrained value for $\epsilon_{\rm cr}$.

For starbursts and Arp 220 without TeV data, we illustrate the
allowed high-energy behavior by plotting the $\pm 1 \sigma$
flattest and steepest curves (the parameters values are the cross points in the contour plots Fig.~\ref{fig:contour}) in addition to the best-fit curves (the parameters values are the central points in Fig.~\ref{fig:contour}).
Comparing these curves at TeV range with the sensitivities of {\em VERITAS}, {\em H.E.S.S} and {\em CTA}, we see that in the optimistic cases,
{\em VERITAS} and {\em H.E.S.S} could measure the TeV signals from NGC 1068, NGC 4945 and the Circinus galaxy.
{\em CTA} should perform well for all the five starbursts, and may be able to detect Arp 220 in a long-term observation as Arp 220's TeV flux is around the sensitivity of {\em CTA} in 50 hours \citep{Hassan 2015}.

\subsection{Calorimetric Limit}
\label{subsec:Calorimetric Limit}

From eq.~\ref{eq:gammalumSFR}, in our closed box model the ratio
of gamma-ray luminosity to the star-formation rate $\psi$
depends only on the $(\ecrsn,s)$ parameters.  Further,
a galaxy's star formation rate $\psi$ scales with its
far IR luminosity due to reprocessing of starlight by dust
\citet{Kennicutt 1998},
\beq
\label{eq:Ken}
\frac{\psi}{M_{\odot}yr^{-1}}= 1.3\times10^{-10}\frac{L_{8-100\mu m}}{L_{\odot}}
\eeq
where the proportionality constant used here \citep{Acker2012}
is for a \citet{Chabrier 2003} initial mass function. 
In closed-box gamma emitters, therefore,
the ratio $L_{\gamma}/L_{\rm IR}$ also only depends on $\ecrsn$ and $s$:
\beqar
\label{eq: gammalumradio}
L_{\gamma}/L_{8-100\mu m}& = &\frac{L_{\gamma}}{\psi} \frac{\psi}{L_{8-100\mu m}}
\eeqar
where ${L_{\gamma}}/{\psi}$ is from our model's eq.~\ref{eq:gammalumSFR}, ${\psi}/{L_{8-100\mu m}}= 1.3\times10^{-10} M_{\odot}yr^{-1}/{L_{\odot}}$ from eq.~\ref{eq:Ken}.
This ratio provides a measure of calorimetry
as we have defined it and encoded in our model.

The expected calorimetric limit ratio $L_{\rm >1GeV}/L_{8-100 \mu m}$ for CR nuclei with $s=2.0$ is $L_{\gamma}/L_{8-100\mu m}\sim 5.2\times10^{-4} (\epsilon_{\rm cr}/0.3 {\ \rm foe})$ for our thick-target model.  For comparison,  this is significantly higher than \citet{Thompson2007}'s $ \sim 10^{-5} (\epsilon_{\rm cr}/0.05 {\ \rm foe})$, but is in good agreement with \citet{Lacki et al. 2011} ratio $3.1\times10^{-4}(\epsilon_{\rm cr}/0.1 {\ \rm foe})$ and with {\em Fermi} group's result $2.5\times10^{-4}(\epsilon_{\rm cr}/0.1 {\ \rm foe})$ \citep{Acker2012}.

The systematic uncertainties of our calorimetric model's gamma-ray luminosity
mainly come from two sources.  One is
the uncertainty in the $L_{\rm IR}$-SN rate conversion.
While the $L_{\rm IR}$-SFR conversion introduces the error with a factor of 2-3 \citep{Kennicutt 1998}, the fact that both SN rate and far-IR luminosity arise from massive stars brings a cancellation of the error, making the final $L_{\rm IR}$-SN rate calibration uncertainty as good as 10-20$\ \rm{percent}$ \citep{Horiuchi 2011}.
The other main uncertainty in our model is the cross section $\sigma_{\rm pp}$ of $p-p$ reaction that is generally better than $10\ \rm{percent}$ \citep{Olive 2014}.
Furthermore, the calorimetric gamma-ray luminosity derives from the ratio
$\sigma_{\rm pp,inelastic}/b(\sigma_{\rm pp,total})$, making additional cancellation of the uncertainty.
So the resultant calorimetric gamma-ray luminosity should be good to $\lesssim30\ \rm{percent}$ or better.

\begin{figure*}
\centering
\includegraphics[width=1.2\columnwidth]{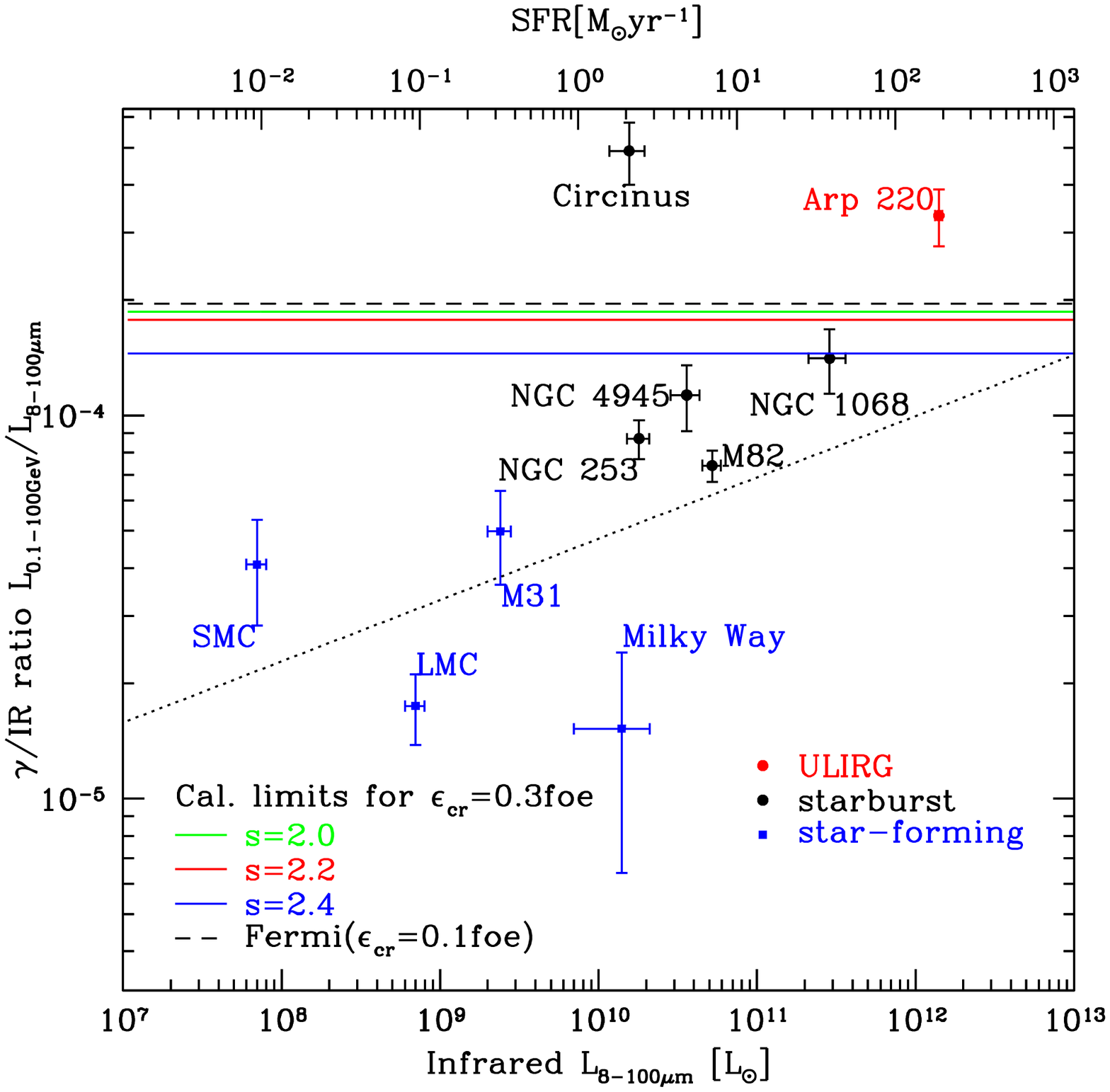}
\caption{\small
Plot of ratio of gamma-ray luminosity ($0.1-100{\rm GeV}$) to total IR luminosity (8-100${\rm \mu m}$). Blue squares: ordinary star-forming galaxies; black points:  starbursts; red: ULIRGs. Milky Way IR and gamma-ray results from \citet{Strong et al. 2010}, IR data for other galaxies from \citet{Sanders 2003}, gamma-ray data for SMC \citep{Abdo et al. 2010b}, LMC \citep{Abdo et al. 2010a}, M31 \citep{Abdo et al. 2010c}).  Starburst IR data from \citet{GS2004}, gamma-ray data from \citet{Acker2012}, except for the Circinus \citep{Hayashida et al. 2013} and Arp 220 \citep{Peng 2016}. The black dotted line: \textit{Fermi}'s best-fit power law relation \citep{Acker2012}.  Upper abscissa: SFR estimated from the IR luminosity \citep{Kennicutt 1998}. The blue solid line:  calorimetric gamma-ray luminosity limit assuming an average CR acceleration energy per supernova of $\epsilon_{\rm cr} =3 \times 10^{50}{\rm erg}=0.3\ {\ \rm foe}$ with source CR index $s=2.4$; purple and green lines for $s=2.2$ and $s=2.0$ respectively.The black dashed line indicated {\em Fermi}'s calorimetric results ($s=2.2,\epsilon_{\rm cr}=10^{50}{\rm erg} $) \citep{Acker2012}.
\label{fig:cal}
}
\end{figure*}

The our limit $L_{\rm 0.1-100 GeV}/L_{\rm 8-100 \mu m}$ is plotted in Fig.~\ref{fig:cal}
for different choices of CR proton index $s$. 
Note that our calorimetric limits agree with {\em Fermi} group's \citep{Acker2012} within $30\ \rm{percent}$,
which is consistent within uncertainties. 

In each of our calculations and plots for individual galaxies,
the cosmic-ray acceleration efficiencies correspond to a mean value for
all supernovae in the galaxy.  We can compare this to typical values of $\epsilon_{\rm cr}$ for Milky
Way supernovae taken from the literature.
These values typically vary \citep[e.g.,][]{Fields et al. 2001}
from 0.1 ${\ \rm foe}$ to 0.3 ${\ \rm foe}$ (see \S~\ref{subsec: Index and SNR}).
We provisionally adopt a {\em maximum} value of $\epsilon_{\rm cr,max}=0.3\ {\rm foe}$ 
in order to judge the proton calorimetry of the starbursts. 
If $\epsilon_{\rm cr}>{\epsilon_{\rm cr,max}}$, calorimetry fails for that galaxy, because our model gives an upper-limit to the gamma-ray spectrum, possible explanations are discussed later in this section; if $\epsilon_{\rm cr}<{\epsilon_{\rm cr,max}}$, the starburst is a proton calorimeter with the calorimetric efficiency 
\beqar
\centering
\eta_{\rm cal}&=&\frac{\gamma \text{-ray derived CR acceleration per SN}}{\text {maximum CR acceleration per SN} }
\nonumber\\
&=&\frac{\epsilon_{\rm cr}}{\epsilon_{\rm cr,max}}
\label{eq:efficiency}
\eeqar
i.e., M82 has a calorimetric efficiency of $35\ \rm{percent}$, NGC 253 is $39\ \rm{percent}$, NGC 1068 is $84\ \rm{percent}$ and NGC 4945 is $70\ \rm{percent}$. For the Circinus galaxy and the ULIRG Arp 220, there are two possibilities: the galaxy is a fully proton calorimeter (the calorimetric efficiency is $100\ \rm{percent}$) with different CR behavior; the calorimetry relation fails. 

The proton calorimetry of the starbursts could also be judged by Fig.~\ref{fig:cal}, which shows both the calorimetric limit from our model and data for all star-forming galaxies with gamma-ray detections.
Here there are two measurements of the ULIRG Arp 220.
\citet{Griffin 2016} measure the luminosity of Arp 220 to be $8.22\pm3.0\times10^{41}{\rm ergs/s}$ in the energy band $[0.8,100] {\rm GeV}$, while our model's calorimetric limit $L_{\gamma}$ in the same energy range is $5.7\times10^{41}{\rm ergs/s}$; another independent group \citet{Peng 2016} report their gamma-ray luminosity to be $1.39\pm0.31\times10^{42}{\rm ergs/s}$ in the energy band $[0.2,100] {\rm GeV}$, while our calorimetric limit result is $0.95\times10^{42}{\rm ergs/s}$. 
Therefore, although Arp 220 is high above the calorimetric limits in Fig.~\ref{fig:cal}, within the errors, the observed gamma-ray luminosity is not far from or even compatible with our model's calorimetric limit in the same energy range. 

Fig.~\ref{fig:cal} allows us to draw several conclusions.
\begin{enumerate}
\item
Normal, Milky-Way-like (``quiescent'') star-forming galaxies are about an order of magnitude below the calorimetric limits.
This is as expected:  Milky-Way Galactic cosmic rays are known to be escape-dominated
and thus their cosmic rays find themselves in the thin-target regime,
rather than thick-target calorimetric limit.  We see that for these systems,
most ($\sim 90\ \rm{percent}$) cosmic rays escape before interacting.

\item
The starburst galaxies M82, NGC 253, NGC 1068 and NGC 4945 are close to the limits, which
shows that calorimetry is a good approximation for these galaxies.  This
further implies that quiescent and starburst galaxies occupy opposite limits of
gamma-ray production.

\item
Two galaxies lie above the calorimetric bounds.
The Circinus galaxy lies substantially above these limits.
For Arp 220, the situation is somewhat less clear.
\end{enumerate}

In the case that a galaxy's gamma-ray emission
truly exceeds our bound on proton calorimetry,
there are several possible explanations.
Two possibilities envision increased pionic emission from cosmic-ray protons, 
so that the galaxy remains fully a proton calorimeter  \citep{Torres 2004, Lacki & Thompson 2013,YH2015}.
This could occur if a galaxy harbors supernovae that are systematically more
efficient accelerators than in the present Milky Way, i.e.,
exceeding our adopted value $\epsilon_{\rm cr,max} = 0.3\ \rm foe$. 
Presumably this would reflect systematically more energetic explosions
and/or more favorable particle injection.  A test for this scenario
would be that cosmic-ray electron signatures should be similarly enhanced, e.g.,
radio synchrotron, or IC emission below the pion bump.
A higher pionic flux would also follow if supernova rates are
underestimated by far-infrared luminosity measurements, i.e, the scaling relation between the far-infrared luminosity and SFR/supernova rate is different \citep[e.g.][]{ Hayashida et al. 2013, Fox 2015}. This would
require that {\em less} UV from massive stars is reprocessed by dust
than in quiescently star-forming galaxies, which seems difficult to arrange in starburst
and/or ULIRGs.

A galaxy may also exceed the calorimetric bound
because the gamma-ray emission is dominated by sources
other than protons \citep[e.g.,][]{Downes & Eckart 2007, Sakamoto 2008, Wilson 2014, Tunnard 2015, YH2017}.
Electron gamma emission could dominate if there is a much larger electron/proton ratio in the galaxy's cosmic rays,
or if proton escape is important (also see \S~\ref{subsec:Assumptions} for primary and secondary electron emissions as well as diffusion and advection loss effects); this would imply that the gamma-ray spectrum should not show a pion feature.
Finally, a galaxy can exceed our bound if it harbors an active
nucleus in which a supermassive black hole jet powers gamma-ray emission.
A signature here would be the time variability that is characteristic
of most gamma-ray signals from active galaxies.

\subsection{Neutrino estimation for individual starbursts}
\label{subsubsec: neutrino}

The same CR-ISM interactions that produce gamma rays also generate cosmic neutrinos,  because $pp$ collisions create both neutral and charged pions \citep[e.g.,][]{Halzen & Hooper 2002}.  The charged pions decay to neutrinos via $\pi^{+}\to \nu_{\mu}\bar{\nu}_{\mu}\nu_{\rm e}e^{+}$ and $\pi^{-}\to \bar{\nu}_{\mu}\nu_{\mu}\bar{\nu}_{\rm e}e^{-}$. Thus starburst galaxies are guaranteed high-energy neutrino sources \citep[e.g.,][]{Loeb & Waxman 2006, Lacki et al. 2011}, though their detectability depends upon the detection sensitivity.

In $pp$ collisions, isospin considerations demand that $N_{\pi^{\pm}}\simeq2N_{\pi^0}$ and the flavor ratio after oscillations is $\nu_{\rm e}:\nu_{\mu}:\nu_{\tau}=1:1:1$ for both neutrinos and antineutrinos \citep{Kamae et al. 2006, Kelner et al. 2006}.  The differential fluxes of gamma-rays and single-flavor neutrino ($\nu_i+\bar{\nu}_i, {\rm i=e,\mu,\tau}$) are approximately related as $dF_{\nu_i}/dE_{\nu_i}(E_{\nu_i} \approx E_{\gamma}/2)=2dF_{\gamma}/dE_{\gamma}(E_{\gamma})$ ignoring kinematic differences and absorption effects \citep{Anchordoqui et al. 2004, Ahlers & Murase 2014, Murase et al. 2013}. Therefore for a given starburst galaxy, we estimate the upper-limit to its neutrino flux at high energy by our model's calorimetric pionic gamma-ray result and thus constrain the flux measured from neutrino telescopes like IceCube. 

For the case of M82, our model gives a flux $F_{\gamma,\rm 2TeV-2PeV}\sim 5.2 \times 10^{-14}{\rm ph cm^{-2} s^{-1}}$, the associated single-flavor neutrino flux (${\rm 1TeV-1PeV}$) would thus be $F_{\nu, \rm 1TeV-1PeV}\sim 1.0 \times 10^{-13}{\rm cm^{-2} s^{-1}}$, $dF_{\nu}/dE_{\nu}\sim 6.6 \times 10^{-14}(E_{\nu}/TeV)^{-2.275}{\rm cm^{-2} s^{-1}}$, $dF_{\nu}/dE_{\nu}(E_{\nu}=1{\rm TeV})\sim 6.6 \times 10^{-14}{\rm cm^{-2} s^{-1}}$.  For IceCube, the median sensitivity at $90\ \rm{percent}$ CL is $\sim{\rm 10^{-12}TeV^{-1}cm^{-2}s^{-1}}$ for energies between ${\rm 1TeV-1PeV}$ with an $E^{-2}$ spectrum and the upper-limit of M82 got by IceCube $\Phi_{\nu_{\mu}+\bar{\nu}_\mu}^{90\ \rm{percent}}=2.94\times 10^{-12}{\rm TeV^{-1}cm^{-2}s^{-1}}$ \citep{IceCube 2014}. Therefore our model's estimated neutrino flux of M82 is well below the upper limit, and is more than $10$ times too faint to be observed by current IceCube, in agreement with \citet{Lacki & Thompson 2013} and \citet{Murase2016}'s conclusion. However, stacking searches of starbursts may get a detectable signal in the next generation detectors \citep{Lacki et al. 2011, Murase2016}, and the starbursts can contribute to the diffuse neutrino background that may also be detectable \citep{Loeb & Waxman 2006}.

\section{Discussion and Conclusions}
\label{sec:Discussion and Conclusions}

We have constructed a two-parameter, closed-box, thick-target model to explain the gamma-ray emission from, and test the cosmic-ray calorimetry of, starburst galaxies.
\citet{Pohl1993, Pohl1994} presented a prescient 
theoretical study of the calorimetric behavior of galaxies in
the EGRET era.   He characterized star-forming galaxies in
the thick-target limit as
``fractional calorimeters'' for both hadrons and leptons.
Specifically, Pohl pointed out that
fraction of cosmic-ray energy returned in gamma-rays reflects a combination of
the fraction of particle loss mechanism that can lead to gammas, and
the branching to gammas in those interactions.
Our approach is guided by this point of view, and we now
have the benefit of GeV and TeV data on star-forming galaxies
to test these ideas.

In addition, gamma-ray emission from starburst galaxies has been calculated by a number of
groups \citep[e.g.,][]{Paglione et al. 1996, Torres 2004, Persic et al. 2008, de Cea del Pozo et al. 2009, Lacki et al. 2010, Lacki et al. 2011, YH2013, Eichmann 2016}. These important papers follow
calculation procedures similar to ours, and also solve the one-zone diffusion-loss equation \citep[e.g.,][]{Mene1971, Longair 1981} to obtain steady-state particle spectrum and in turn the gamma-ray emission.
However, these papers and ours differ in several assumptions, variables and formula numerical calculations.  
 (1) Previous treatments use the general solution to the diffusion-loss equation with different parameter choices, except for \citep{YH2013,YH2014,YH2015} who adopt an approximate solution with loss dominant assumption and diffusion emitted, while ours is a thick target approximation with a ``closed-box'' calculation, restricting ourselves to proton interactions in order to place a firm and well-defined upper-limit of the hadronic gamma emission. 
 (2) In order to get the pionic gamma-ray spectrum $dq_{\gamma}/dE_{\gamma}$ (eq.~\ref{eq:gammaspectrum}), we carry a full numerical evaluation of the emissivity $dq_{\pi}/dE_{\pi}$ (eq.~\ref{eq:pionspectrum}), while other groups either adopt the parameterization equations of differential cross section $d \sigma_{\pi}(E_p,E_{\pi})/dE_{\pi}$ directly \citep[e.g.,][]{Torres 2004, Domingo & Torres 2005}, or use $\, {\sc GALPROP}$ code to calculate the differential cross section from pp collision \citep[e.g.,][]{Lacki et al. 2010}, or assume a delta function approximation for pion distribution \citep{YH2013,YH2014,YH2015}, or directly use the analytical form of the gamma energy distribution given by \citet{Kelner et al. 2006} \citep{Eichmann 2016}. 
(3) These calculations to various extents present multi-frequency
and multi-process models, i.e., radio plus gamma-ray emission, with both leptonic process (synchrotron, bremsstrahlung and inverse Compton) and hadronic process (pion decay) in a more complex and realistic way. This
naturally entails more free parameters like the source CR parameters for both electrons and protons, diffusion loss time scale, advection loss timescale, magnetic field, gas density.

Our  model {\em by construction} is less
ambitious than these other studies, but very well-defined with only two parameters ($s,\epsilon_{\rm cr}$).
Our results are, for example, independent of the galaxy's gas density.
Thus our model is targeted to (1)
offer a particularly direct and simple means of estimating
these fundamental parameters in starburst galaxies,
thus measuring their cosmic-ray acceleration properties
that can be compared with those in the Milky Way;
and (2) 
place a firm and careful upper limit to the hadronic gamma-ray luminosity
of {\em any} star-forming galaxy.

For individual starburst galaxies, our model gives good fits to the gamma-ray data in both GeV and TeV range with proper choices of the injected proton index $s$ and cosmic-ray proton acceleration energy per supernova $\epsilon_{\rm cr}$, showing the thick-target assumption is a plausible explanation of the observed starburst GeV and TeV emission.
Our model shows that the gamma-ray spectrum of thick-target systems
shares the same index as the CR ``injection'' index,  instead of the CR propagated index.  This contrasts with the ``thin-target'' situation that should correspond to ordinary star-forming galaxies like Milky Way.
Our fit gives the average value of $s$ in starbursts to be $\sim2.3$, which is consistent with the LAT measurement of Galactic SNRs with an average value of $s$ to be 2.39 \citep{Acero et al. 2016},
implying that cosmic-ray acceleration by supernovae is broadly similar
in starburst galaxies and the Milky Way.  

The goodness of our fit of starbursts M82, NGC 253, NGC 1068 and NGC 4945 suggest that starburst galaxies are proton calorimeters with calorimetric efficiencies vary from $35\ \rm{percent}$ to $84\ \rm{percent}$. These efficiencies may be different in reality if the actual supernova acceleration of CR rate in starbursts differ from the maximum CR acceleration energy $\epsilon_{\rm cr,max}=0.3\ \rm foe$ we have adopted; the scaling is simply $\eta_{\rm cal}=\epsilon_{\rm cr}/\epsilon_{\rm cr,max}$ (eq.~\ref{eq:efficiency}) . For the Circinus galaxy, our model's gamma-ray luminosity agrees with \citet{Hayashida et al. 2013}, and
is above our limit, as is the ULIRG Arp 220. The gamma excesses may be explained in two ways: the galaxy is a full proton calorimeter or proton calorimetry fails for the galaxy, detailed discussions see \S\ref{subsec:Calorimetric Limit}.
 Therefore we conclude that at least for currently observed starbursts, most are nearly or fully proton calorimeters. Others have also addressed the
 question of proton calorimetry in starbursts. For example, \citet{YH2013,YH2014} find M82 and NGC 253 50$\ \rm{percent}$ proton calorimeters, \citet{Acker2012} get calorimetric efficiencies of $30\ \rm{percent}-50\ \rm{percent}$ for starburst galaxies with SFR $\sim 10 {\rm M_{\odot}yr^{-1}}$, while \citet{Lacki et al. 2010, Lacki et al. 2011} conclude that proton calorimetry holds for starburst galaxies with
 $\Sigma_{\rm gas}>1{\rm g \ cm^{-2}}$ and the calorimetric fraction is 0.2 for NGC 253 and 0.4 for M82. Moreover, \citet{Torres 2004}, \citet{Lacki & Thompson 2013} and \citet{YH2015} conclude that Arp 220 is a hadronic calorimeter
or nearly so.  Our conclusions are consistent with these.

More data can further test starburst proton calorimetry.
There are no published starburst data at energies $\sim$30-100 {\rm MeV};
observations in this regime should reveal the characteristic ``pion bump.''
TeV data for NGC 1068, NGC 4945, Circinus, and Arp 220 is also needed to constrain the choices of parameters (both $s$ and $\epsilon_{\rm cr}$) in our model with smaller uncertainty. If Arp 220 indeed saturates the proton calorimeter limit,
it is the best example of a star-forming galaxy as a proton calorimeter,
but it lies at the edge of GeV detectability and has no TeV measurements.
 As discussed in \S\ref{subsec:individual starbursts}, {\em VERITAS} or {\em H.E.S.S} could measure the TeV signals from the starbursts NGC 1068, NGC 4945 and the Circinus galaxy within their sensitivities. Future {\em CTA} observations should dramatically improve
our understanding of starburst galaxies, and may be able to detect Arp 220 in a long-term dedicated observation.

There still remains space to improve our model. Future work would benefit from better 
observational determination of galaxy distances, star-formation and supernova
rates, and of course well-measured TeV gamma-ray data. 
The particle experimental data adopted in our model is as old as from 1980s, we would like to call for new measurements of the pion momentum distribution in
the $p-p$. These data are important not only for gamma-ray emissions but also for the inelastic losses of CRs.
Theoretical work would benefit from additional multi-wavelength constrains on the cosmic-ray electrons (add leptonic process in our model).  Finally, if a starburst could be resolved spatially, perhaps in the TeV,
this would motivate consideration of the supernova and gas distributions inside a starburst.

\section*{Acknowledgments}

We are pleased to thank Keith Bechtol for providing the Fermi data points, Wystan Benbow for providing VERITAS data points, Roger Blandford, Ellen Zweibel and Anne Sickles for the stimulating conversation. This work was supported in part by the NASA Astrophysics Theory Program 
through award NNX10AC86G. We also thank the referee for her/his productive and valuable comments.

\appendix

\section{Order-Of-Magnitude Estimates}
\label{sec:Order-Of-Magnitude Estimation}

An order of magnitude calculation of our model will help to give a sense of the final results and frame key physical issues. We aim to find the calorimetric gamma-ray emission from individual starburst galaxies.

For a starburst galaxy, the injected cosmic-ray energy rate got from supernovae exploration is:
\beqar
\label{eq:crluminosity}
dE_{\rm cr}/dt & = &f_{\rm cr} E_{\rm sn} R_{\rm sn}=\epsilon_{\rm cr}R_{\rm sn}
\nonumber\\
& = & \int E_{\rm p}dN_{\rm p}/dt=L_{\rm cr}
\eeqar
assuming the injected cosmic-ray spectrum is a power law in momentum here, $q_{\rm p}=dN_{\rm p}/dE_{\rm p}dt=dq/dp_{\rm p}=C p_{\rm p}^{-s}, dE_{\rm cr}/dt = C \int_{p_{\rm min}} E_{\rm p} p_{\rm p}^{-s}dp_{\rm p}, dN_{\rm cr}/dt =C\int_{p_{\rm min}}p_{\rm p}^{-s}dp_{\rm p}$, where C is a constant, $p_{\rm min}$ is the minimal momentum of injected CR protons that can be accelerated by SN.

Our model assumes all cosmic-rays will interact with interstellar medium, the interactions involve both elastic and inelastic scattering, in the GeV energy range. Thus we can get a crude estimation that the elastic scattering CR number is about the same as the inelastic number, i.e., $dN_{\rm cr, inelastic}/dt \sim dN_{\rm cr, elastic}/dt \sim (dN_{\rm cr}/dt)/2$.
For the inelastic scattering, only neutral pions could decay into photons, which take up one third of the total produced pion numbers, therefore $dN_{\gamma}/dt=2dN_{\pi^{0}}/dt\sim 2(dN_{\rm cr, inelastic}/dt(E_{\rm cr}>E_{\rm threshold}))/3\sim (dN_{\rm cr}/dt(E_{\rm cr}>E_{\rm threshold}))/3=\dot{N}_{\rm cr, threshold}$, where $E_{\rm threshold}$ is the threshold kinetic energy of CR proton that can produce a pion.

\beqar
\label{eq:I1}
\dot{N}_{\rm cr, threshold}& = &dN_{\rm cr}/dt(E_{\rm cr}>E_{\rm threshold})
\nonumber\\
& \propto & \int_{p_{\rm threshold}}p_{\rm p}^{-s}dp_{\rm p}\propto \frac{p_{\rm threshold}^{1-s}}{s-1}
\eeqar

In this case, we can get an estimation of the gamma-ray (number) flux from the thick-target model is:
\beqar
\label{eq:gammafluxe}
F_{\gamma} & = & \frac{1}{4\pi d^2}\frac{dN_{\gamma}}{dt} 
\nonumber\\
 & = & \frac{1}{4\pi d^2} \frac{dN_{\gamma}/dt}{dE_{\rm cr}/dt} dE_{\rm cr}/dt
\nonumber\\
& = & \frac{1}{4\pi d^2} \epsilon_{\rm cr} R_{\rm sn} \frac{1}{3} f_{\rm threshold}
\eeqar
where $f_{\rm threshold}=\dot{N}_{\rm cr, threshold}/L_{\rm cr}$ is the average CR injected energy per above-threshold proton. 

If $p_{\rm p}<m_{\rm p}$, protons can be approximated to be nonrelativistic, thus $E_{\rm p}\approx p_{\rm p}^2/2m_{\rm p}$, while if $p_{\rm p}>m_{\rm p}$, protons can be approximated to be relativistic, $E_{\rm p}\approx p_{\rm p}$, and $2<s<3$, therefore we have:
\beqar
\label{eq:I2}
L_{\rm CR} & \propto & \int_{p_{\rm min}} E_{\rm p}p_{\rm p}^{-s}dp_{\rm p}
\nonumber\\
& \propto & \int_{p_{\rm min}}^{m_{\rm p}}\frac{p_{\rm p}^2}{2m_{\rm p}}p_{\rm p}^{-s}dp_{\rm p}+\int_{m_{\rm p}}^{\infty}p_{\rm p}^{1-s}dp_{\rm p}
\nonumber\\
&\propto& \frac{m_{\rm p}^{2-s}}{2(3-s)}[1+\frac{6-2s}{s-2}-(\frac{p_{\rm min}}{m_{\rm p}})^{3-s}]
\eeqar
For fixed $s$, $\delta L_{\rm cr} \sim (\delta p_{\rm min}/m_{\rm p})^{3-s} \sim (\delta E_{\rm min}/m_{\rm p})^{\frac{3-s}{2}}$, when s=2.2, $\delta L_{\rm cr} \sim (\delta E_{\rm min}/m_{\rm p})^{0.4}$.Therefore for $2<s<3$, we can see that $L_{\rm cr}$ from CR spectrum is insensitive to $p_{\rm min}$, which is fortunate as there is no accurate determination of $p_{\rm min}$, and most $L_{\rm cr}$ comes from $p_{\rm p}\sim m_{\rm p}$.

Let $s=2.2, f_{\rm cr}=0.1, E_{\rm sn}=10^{51}{\rm erg}, \epsilon_{\rm cr}=10^{50}{\rm erg}, E_{\rm min}=0.001{\rm GeV}, E_{\rm threshold}=0.28{\rm GeV}$,the estimated gamma-ray flux for a certain starburst galaxy with the distance d and supernova rate $R_{\rm sn}$ is $F_{\gamma}\approx 3.31 \times 10^{50} R_{\rm sn}/d^2$.
For the starburst galaxy NGC 253, our oder of magnitude estimation gives the flux to be $4.57\times 10^{-9}{\rm cm}^{-2}{\rm s}^{-1}$, agrees with {\em Fermi} measurement $10.7\pm2.1 \times 10^{-9}{\rm cm}^{-2}{\rm s}^{-1}$ \citep{Hayashida et al. 2013} in an order of magnitude.

\section{Energy Loss Rates}
\label{App:AppendixA}

The energy losses other than Pionic process in our model are elastic scattering and ionization, they are expressed as follows \citep{Gould 1982, Ginzburg & Syrovatskii 1964}:
\beqar
\label{eq:be}
b_{\rm elastic,p} &\sim& 2.44\times10^{-16}\frac{n_{\rm p}}{cm^{-3}}\frac{E_{\rm p}}{\rm GeV}(\frac{E_{\rm p}}{m_{\rm p}c^2})^{1/2}
\nonumber\\
& & \frac{(1+E_{\rm p}/{2m_{\rm p}c^2})^{1/2}}{1+E_{\rm p}/m_{\rm p}c^2} {\rm GeV} {\rm s}^{-1}
\eeqar
\beqar
\label{eq:bi}
b_{\rm ionic,p} &\sim& 1.83\times10^{-17}(\frac{n_{\rm H}+2n_{\rm H_2}}{cm^{-3}})\frac{c}{v_{\rm p}}\{10.9+2\ln(\frac{E_{\rm p}}{m_{\rm p}c^2})
\nonumber\\
& & + \ln(\frac{{v_{\rm p}}^2}{c^2})-\frac{{v_{\rm p}}^2}{c^2}\}{\rm GeV} {\rm s}^{-1}
\eeqar
where $n_{\rm p}$ and $n_{\rm H}+2n_{\rm H_2}$ are the number densities of protons in the ISM, which are equal to $n_{\rm gas}$. Here, $E_{\rm p}$ is the total energy of a proton, $T_{\rm p}$ denotes kinetic energy of a proton. In GeV energy range, elastic scattering contributes about $50\ \rm{percent}$ lower than inelastic scattering does to the total energy-loss during CR propagation. Therefore it is necessary to include elastic scattering during the propagation.

\begin{figure}
\centering
\includegraphics[width=\columnwidth]{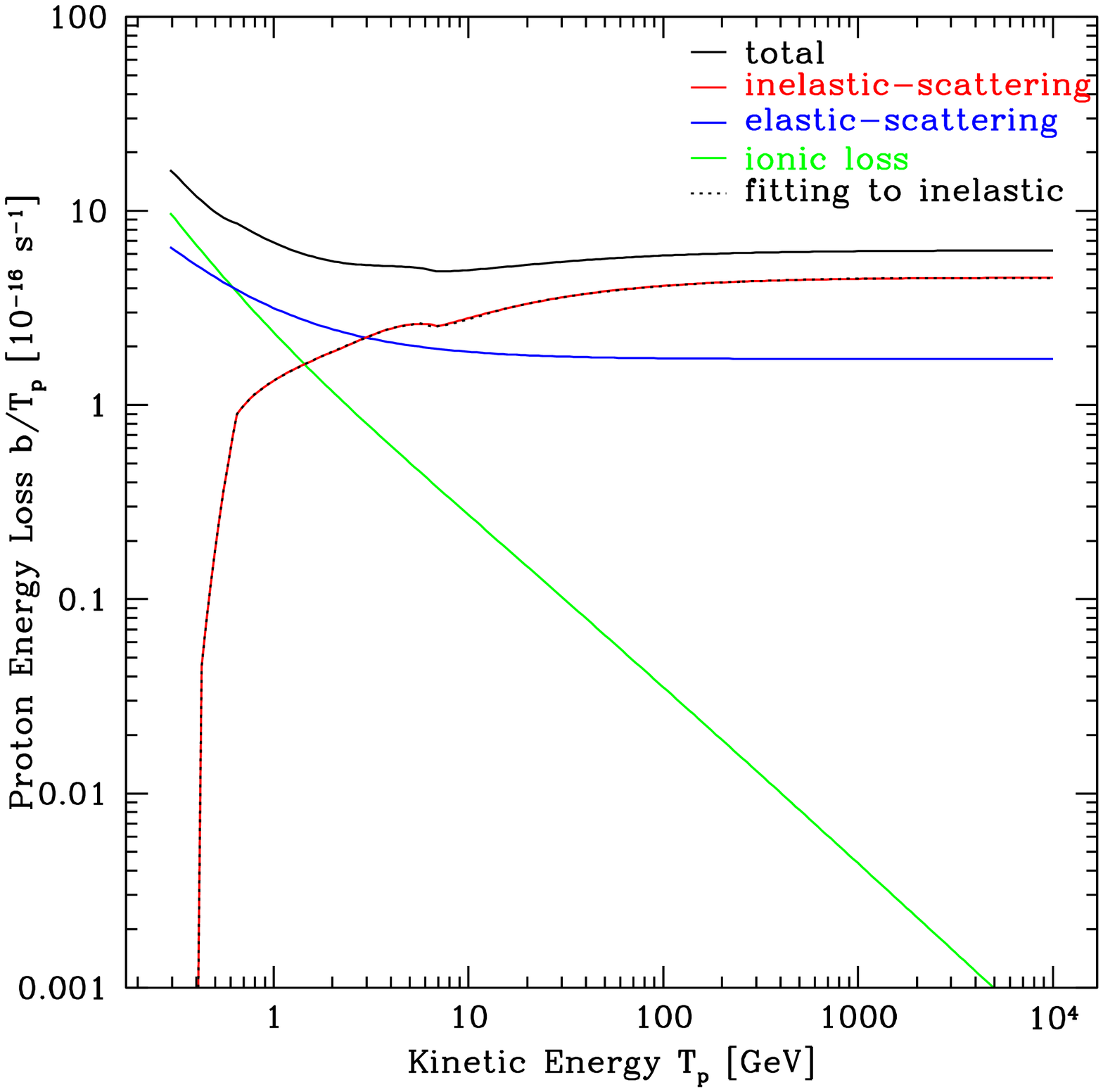}
\caption{\small
Proton Energy Losses. The black line is the total energy loss rate per proton kinetic energy, blue line is elastic energy loss rate per proton kinetic energy, green line is ionic energy loss rate per proton kinetic energy, red line is inelastic (pionic) energy loss rate per proton kinetic energy, black dotted line is our fit curve to inelastic energy loss. Here $n_{\rm gas}=1\ {\rm {cm^{-3}}}$.
\label{fig:b_e}
}
\end{figure}

At high energy, $E_{\rm p}\sim T_{\rm p}$:
As Fig.~\ref{fig:b_e} shows, for $T_{\rm p}>100 {\rm GeV}$, $b(E_{\rm p})\propto E_{\rm p}$, therefore, eq.~\ref{eq:protonspectrum} gives $\phi_{\rm p}\propto {E_{\rm p}}^{-s}$;and for high $T_{\pi}$,$d\sigma_{\pi}(T_{\rm p},T_{\pi})/dT_{\pi}=\avg{\zeta\sigma_{\pi}(T_{\rm p})}dN(T_{\rm p},T_{\pi})/dT_{\pi}\propto 1/T_{\rm p}$, we can get $q_{\pi}\propto E_{\pi}^{-s}$ from eq.~\ref{eq:pionspectrum} in \S~\ref{subsec:Pionic Emission From Individual Starburst Galaxies}, thus $q_{\gamma}\propto E_{\gamma}^{-s}$, or $F_{\gamma}\propto E_{\gamma}^{-s}$.
 Therefore the gamma-ray spectrum obtained from our thick-target model has the same spectral index $s$ as the injected proton's.

An analytical fit to our self-consistent inelastic energy loss appears is shown in Fig.~\ref{fig:b_e} as the black dotted curve. The fit is good with fractional error less than $2\ \rm{percent}$ over the {\em Fermi} energy range. The fitting function is:
\beqar
\label{eq:fit}
Y & = & 0.631x^2+0.502x-0.441,   \textrm{$x_{\rm threshold} \le  x \le -0.24$}
\nonumber\\
  &   &-1.66x^2-0.605x-0.575,    \textrm{$-0.24 \le x \le -0.05$}
\nonumber\\
  &   &-0.430x-0.568,            \textrm{$-0.05 \le x \le 0.6$}
\nonumber\\
  &   &-0.643x-0.440,            \textrm{$0.6 \le x \le 0.75$}
\nonumber\\
  &   &-0.157\ln(x-0.639)-1.26,  \textrm{$0.75 \le x \le 1.10$}
\nonumber\\
  &   &-0.677\ln(x+0.817)-0.701,  \textrm{$1.10 \le x $}
\eeqar
where $Y=\log_{10}(b_{\rm inelastic}/\avg{\zeta \sigma_{\pi}(X)})$ with $x=\log_{10}(T_{\rm p}/1{\rm GeV}), x_{\rm threshold}=\log_{10}(T^{min}_{\rm p}/1{\rm GeV})$, for $n_{\rm gas}=1\ { \rm cm^{-3}}$.

Finally, can use these results to compare collisional timescales to
the timescales for other cosmic-ray losses.  For a starburst, at GeV energy range, the diffusion timescale is
$\tau_{\rm diff} \sim H^2/2D \sim 5\times10^{6} {\rm yr}$, where $H\sim1 {\rm kpc}$ is the height of the disk,
and we use the diffusion coefficient $D\sim3\times 10^{28} {\rm cm^2/s}$ for 1 ${\rm GeV}$ protons in our Galaxy. \footnote{As $D\sim E^{\delta}$ with $\delta\sim0.5$, the escape timescale at TeV will be shorter, but most of the CR energy is around 1 GeV, so escape has little affect on the energy loss for the protons of interest to us. We thank the referee for pointing this out.}
The advective escape timescale $\tau_{\rm adv} \sim r_{\rm s}/v_{\rm wind}\sim 10^6 {\rm yr}$ is the
time for a wind of speed $v_{\rm wind}\sim 300 {\rm km/s}$  to cross the starburst nucleus region of radius $r_{\rm s}\sim 0.3 {\rm kpc}$ \citep{Rephaeli2013,YH2013,Lacki & Thompson 2013}.
The CR interaction loss timescale is $\tau_{\rm loss} \sim E_{\gamma}/b \sim 1\times 10^5 {\rm yr}$ with the atomic hydrogen density of the interstellar medium $n_{\rm gas} \sim 500 {\rm cm}^{-3}$, where $b$ is the rate of energy loss (see Fig.~\ref{fig:b_e} for $E_{\gamma}/b$ value). 

\section{Code Description}
\label{App:AppendixB}

We build a simple code following the calculation in \S~\ref{sec:The Thick-Target/Calorimetric Model}, using the Simpson method to do integration and the relative errors for the integrations set to be $10^{-4}$. Because the model is closed box, we can do conservation check of the code: $N_{\gamma}=2N_{\pi}=2N_{\rm p}/3$ (number conservation), and $L_{\gamma}=L_{\pi}<L_{\rm p}/3$ (energy conservation) \citep{Kelner et al. 2006}. The code results we get fulfill the conservation check. To reduce the CPU time taken for code running, instead of doing the 3-layer integration, we do the first 2-layer integration first to get the values of $q_{\pi}$ vs. $ E_{\pi}$ and store them as vectors, then doing the third integration to get $q_{\gamma}$ simply by doing interpolation and extrapolation to the stored values of $q_{\pi}$.

\section{Nuclear Enhancement Factor}
\label{App:AppendixC}

In the thick-target model, the gamma-ray luminosity follows from
the production and decay of neutral pions, which
are dominantly produced in collisions between cosmic-ray
protons and ISM protons.
Heavier nuclei in both cosmic rays and the ISM an also produce neutral pions.
This effect is encoded in a ``nuclear enhancement factor'' ${\cal A}$ to be multiplied to the gamma-ray yield assuming cosmic-ray protons on ISM protons only: $dq_{\gamma}^{\rm total}/dE_{\gamma}={\cal A}dq_{\gamma}^{\rm pp,only}/dE_{\gamma}$.

Assume all cosmic-ray species ($j={\rm p}$, He, CNO, NeMgSiS, Fe)
have source spectra with the same shape in energy per nucleon
$\epsilon=E_{\rm i}/A_{\rm i}$, and differ only by
cosmic-ray source abundances $y_j^{\rm cr}$:
\beq
\label{eq:yi}
\frac{dq_{j}}{d\epsilon} = y_{j}^{\rm cr} \ \frac{dq_{\rm p}}{d\epsilon}
 \\.
\eeq
Thus the cosmic-ray power needed to accelerate species $j$ is
\beq
\label{eq:li}
L_{{\rm cr,}j} =
\int dV\int E_{j}\frac{dq_{j}}{d\epsilon}d\epsilon
=A_{j} y_{j}^{\rm cr} \ L_{\rm cr,p}
\eeq
and thus the total cosmic-ray source luminosity
scales with the proton luminosity as
\beq
L_{\rm cr} = L_{{\rm cr,p}} \sum_j A_j \ycr_j=\epsilon_{\rm cr}R_{\rm sn}
\eeq
and 
\beq
 q_{\rm cr}=q_{\rm p}\sum_j A_j \ycr_j
\eeq

For a closed-box model, the total flux in species $j$ is
\beqar
\label{eq:phii}
\Phi_{j}=\frac{v_{j}}{b_{j}}\int \frac{dq_{j}}{d\epsilon}d\epsilon
=y_{j}^{\rm cr} \frac{b_{\rm p}}{b_{j}}\Phi_{\rm p}
\eeqar
For energy losses due to nuclear interactions between CR nuclei j and ISM nuclei i, we 
assume that the cross sections for $j+i \rightarrow \pi^0 + \cdots$
scale with the $pp \rightarrow \pi^0$ cross sections as
${\sigma_{\rm ji}^{\rm inelastic}}/{\sigma_{\rm pp}^{\rm inelastic}}
= {\sigma_{\rm ji}^{\rm elastic}}/{\sigma_{\rm pp}^{\rm elastic}}
={\sigma_{\rm ji}^{\rm total}}/{\sigma_{\rm pp}^{\rm total}}$. 
This leads to energy loss rates (per nucleon) for species $j$ of
\beqar
\label{eq:tbi}
b_{j}& = & b_{j}^{\rm inelastic}+b_{j}^{\rm elastic}=b_{\rm pp}\sum\limits_{\rm i}y_{\rm i}^{\rm ISM}\frac{\sigma_{\rm ji}^{\rm t}}{\sigma_{\rm pp}^{\rm t}}
\\
b_{\rm p}& = & b_{\rm pp}\sum\limits_{\rm i}y_{\rm i}^{\rm ISM}\frac{\sigma_{\rm pi}^{\rm t}}{\sigma_{\rm pp}^{\rm t}}
\eeqar
where $y_{\rm i}^{\rm ISM}=n_{\rm i}/n_{\rm p}$.

The emissivities of pions and gamma-ray induced by CR interactions are
\beqar
\label{eq:}
\frac{dq_{\pi}^{\rm ji}}{dE_{\pi}}=\int d\epsilon_{\rm j}n_{\rm i}\Phi_{\rm j}\avg{\zeta_{\rm ji}^{\pi}\sigma_{\rm ji}^{\pi}}\frac{dN}{dE_{\pi}}(\epsilon_{\rm i},E_{\pi})
\eeqar
\beqar
\label{eq:qij}
\frac{dq_\gamma^{ji}}{dE_\gamma}=2\int_{u(E_\gamma)}\frac{dE_{\pi}}{p_{\pi}}\frac{dq_{\pi}^{ji}}{dE_{\pi}}=\int q_{\pi}^{ji}
\eeqar
therefore the nuclear enhancement factor ${\cal A}$ can be expressed as 
\beqar
\label{eq:qte}
{\cal A}& = & \frac{dq_{\gamma}^{\rm total}/dE_{\gamma}}{dq_{\gamma}^{\rm pp,only}/dE_{\gamma}}
\nonumber\\
& = & \int \frac{q_{\pi}^{\rm pp}}{q_{\pi}^{\rm pp,only}}\sum\limits_{j,i}\frac{dq_{\pi}^{\rm ji}/dE_{\pi}}{dq_{\pi}^{\rm pp}/dE_{\pi}}
\nonumber\\
& = & \frac{q_{\rm p}}{q_{\rm cr}}\frac{b_{\rm pp}}{b_{\rm p}}\sum\limits_{j,i}\frac{dq_{\pi}^{\rm ji}/dE_{\pi}}{dq_{\pi}^{\rm pp}/dE_{\pi}}
\eeqar

As discussed in \citet{Abbott 1992,Miller 2007}, the total multiplicity $R_{\rm ji}^{\pi^{0}}$ for making $\pi^{0}$ through collision of nuclei $i+j$ is almost universal, i.e., $R_{\rm ji}^{\pi^{0}}=\avg{\zeta_{\rm ji}^{\pi}\sigma_{\rm ji}^{\pi}}/{\sigma_{\rm ji}^{\rm inelastic}}=R_{\rm pp}^{\pi}=$constant, thus it is safe to assume ${\avg{\zeta_{\rm ji}^{\pi}\sigma_{\rm ji}^{\pi}}}/{\avg{\zeta_{\rm pp}^{\pi}\sigma_{\rm pp}^{\pi}}}={\sigma_{\rm ji}^{\rm t}}/{\sigma_{\rm pp}^{\rm t}}$ for all energy per nucleon $\epsilon$ for CR nuclei j interact with ISM nuclei i, then we can get
\beqar
\label{eq:A1}
{\cal A}& = & \frac{q_{\rm p}}{q_{\rm cr}}\sum\limits_{j}y_{\rm j}^{\rm cr}
\nonumber\\
&= &\frac{1}{\sum\limits_{j}A_{\rm j}y_{\rm j}^{\rm cr}} \sum\limits_{j}y_{\rm j}^{\rm cr}=\frac{1}{\avg{A}_{\rm cr}}.
\eeqar

Considering the same heavier nuclei components in both CR and ISM as \citet{Mori 2009} did, if use the relative abundance of ${\rm H:He:CNO:NeMgSiS:Fe}$=1:0.153:$1.245\times 10^{-2}$:$3.65\times 10^{-3}$:$1.182\times 10^{-3}$ in \citet{Meyer 1985}, the nuclear enhancement factor is ${\cal A}=0.59$.

\label{lastpage}

\begin{thebibliography}{99}

\bibliographystyle{alpha}


\bibitem[Abbott et al.(1992)]{Abbott 1992} Abbott, T., Akiba, Y., Beavis, D., et al.\ 1992, \prd, 45, 3906 

\bibitem[Abdo et al.(2009)]{Abdo et al. 2009} Abdo, A.~A., Ackermann, M., Ajello, M., et al.\ 2009, Physical Review Letters, 103, 251101

\bibitem[Abdo et al.(2010a)]{Abdo et al. 2010a} Abdo, A.~A., Ackermann, M., Ajello, M., et al.\ 2010, \aap, 512, A7

\bibitem[Abdo et al.(2010b)]{Abdo et al. 2010b} Abdo, A.~A., Ackermann, M., Ajello, M., et al.\ 2010, \aap, 523, A46

\bibitem[Abdo et al.(2010c)]{Abdo et al. 2010c} Abdo, A.~A., Ackermann, M., Ajello, M., et al.\ 2010, \aap, 523, L2

\bibitem[Abdo et al.(2010d)]{Abdo et al. 2010d} Abdo, A.~A., Ackermann, M., Ajello, M., et al.\ 2010, \apjl, 709, L152




\bibitem[Abramowski et al.(2012)]{Abramowski et al. 2012} Abramowski, A., Acero, F., Aharonian, F., et al.\ 2012, \apj, 757, 158

\bibitem[Acciari et al.(2009)]{VERITAS Collaboration et al. 2009} Acciari, V.~A., Aliu, E., Arlen, T., et al.\ 2009, \nat, 462, 770

\bibitem[Acero et al.(2009)]{Acero et al. 2009} Acero, F., Aharonian, F., Akhperjanian, A.~G., et al.\ 2009, Science, 326, 1080

\bibitem[Acero et al.(2016)]{Acero et al. 2016} Acero, F., Ackermann, M., Ajello, M., et al.\ 2016, \apjs, 224, 8 

\bibitem[Ackermann et al.(2012)]{Acker2012} Ackermann, M., Ajello, M., Allafort, A., et al.\ 2012, \apj, 755, 164 

\bibitem[Ackermann et al.(2013)]{Ackermann et al. 2013} Ackermann, M., Ajello, M., Allafort, A., et al.\ 2013, Science, 339, 807 

\bibitem[Aharonian et al.(2005)]{Ah2005} Aharonian, F., Akhperjanian, A.~G., Bazer-Bachi, A.~R., et al.\ 2005, \aap, 441, 465

\bibitem[Aharonian et al.(2008)]{Aharonian et al. 2008} Aharonian, F., Akhperjanian, A.~G., Bazer-Bachi, A.~R., et al.\ 2008, \aap, 481, 401

\bibitem[Ahlers \& Murase(2014)]{Ahlers & Murase 2014} Ahlers, M., \& Murase, K.\ 2014, \prd, 90, 023010  

\bibitem[Aliu et al.(2013)]{Aliu et al. 2013} Aliu, E., Archambault, S., Arlen, T., et al.\ 2013, \apj, 764, 38

\bibitem[Anchordoqui et al.(2004)]{Anchordoqui et al. 2004} Anchordoqui, L.~A., Goldberg, H., Halzen, F., \& Weiler, T.~J.\ 2004, Physics Letters B, 600, 202

\bibitem[Baade 
\& Zwicky(1934)]{Zwicky} Baade, W., \& Zwicky, F.\ 1934, Proceedings of the National Academy of Science, 20, 259

\bibitem[Bell(1978)]{Bell1978} Bell, A.~R.\ 1978, \mnras, 182, 147

\bibitem[Blandford \& Ostriker(1978)]{Blandford1978} Blandford, R.~D., \& Ostriker, J.~P.\ 1978, \apjl, 221, L29

\bibitem[Blasi \& Amato(2012)]{Blasi & Amato 2012} Blasi, P., \& Amato, E.\ 2012, \jcap, 1, 10 

\bibitem[Blom et al.(1999)]{Blom et al. 1999} Blom, J.~J., Paglione, T.~A.~D., \& Carrami{\~n}ana, A.\ 1999, \apj, 516, 744

\bibitem[Caprioli(2012)]{Caprioli 2012} Caprioli, D.\ 2012, \jcap, 7, 38

\bibitem[Chabrier(2003)]{Chabrier 2003} Chabrier, G.\ 2003, \pasp, 115, 763

\bibitem[de Cea del Pozo et al.(2009)]{de Cea del Pozo et al. 2009} de Cea del 
Pozo, E., Torres, D.~F., \& Rodriguez Marrero, A.~Y.\ 2009, \apj, 698, 1054 

\bibitem[Dermer(1986)]{Dermer 1986} Dermer, C.~D.\ 1986, \aap, 157, 223

\bibitem[Dermer \& Powale(2013)]{Dermer & Powale 2013} Dermer, C.~D., \& Powale, G.\ 2013, \aap, 553, A34

\bibitem[Domingo-Santamar{\'{\i}}a \& Torres(2005)]{Domingo & Torres 2005} Domingo-Santamar{\'{\i}}a, E., \& Torres, D.~F.\ 2005, \aap, 444, 403 

\bibitem[Downes \& Eckart(2007)]{Downes & Eckart 2007} Downes, D., \& Eckart, A.\ 2007, \aap, 468, L57 

\bibitem[Eichmann \& Becker Tjus(2016)]{Eichmann 2016} Eichmann, B., \& Becker Tjus, J.\ 2016, \apj, 821, 87 

\bibitem[Evoli et al.(2008)]{Evoli et al. 2008} Evoli, C., Gaggero, D., Grasso, D., \& Maccione, L.\ 2008, \jcap, 10, 18 

\bibitem[Fermi Collaboration (2013)]{Dermer et al. 2013} Fermi Collaboration, 2013, proc. ICRC, p. 1165

\bibitem[Fields et al.(1994)]{Fields 1994} Fields, B.~D., Olive, K.~A., \& Schramm, D.~N.\ 1994, \apj, 435, 185 

\bibitem[Fields et al.(2001)]{Fields et al. 2001} Fields, B.~D., Olive, K.~A., Cass{\'e}, M., \& Vangioni-Flam, E.\ 2001, \aap, 370, 623

\bibitem[Fields et al.(2010)]{Fields et al. 2010} Fields, B.~D., Pavlidou, V., \& Prodanovi{\'c}, T.\ 2010, \apjl, 722, L199 

\bibitem[Fox \& Casper(2015)]{Fox 2015} Fox, O.~D., \& Casper, C.\ 2015, IAU General Assembly, 22, 2258045

\bibitem[Gao \& Solomon(2004)]{GS2004} Gao, Y., \& Solomon, P.~M.\ 2004, \apj, 606, 271

\bibitem[Ginzburg \& Syrovatskii(1964)]{Ginzburg & Syrovatskii 1964} Ginzburg, V.~L., \& Syrovatskii, S.~I.\ 1964, The Origin of Cosmic Rays, New York: Macmillan, 1964

\bibitem[Gould(1982)]{Gould 1982} Gould, R.~J.\ 1982, \apj, 263, 879

\bibitem[Griffin et al.(2016)]{Griffin 2016} Griffin, R.~D., Dai, X., \& Thompson, T.~A.\ 2016, \apjl, 823, L17 

\bibitem[Halzen \& Hooper(2002)]{Halzen & Hooper 2002} Halzen, F., \& Hooper, D.\ 2002, Reports on Progress in Physics, 65, 1025

\bibitem[Hassan et al.(2015)]{Hassan 2015} Hassan, T., Arrabito, L., Bernl{\"o}r, K., et al.\ 2015, preprint (arXiv:1508.06075) 

\bibitem[Hayashida et al.(2013)]{Hayashida et al. 2013} Hayashida, M., Stawarz, {\L}., Cheung, C.~C., et al.\ 2013, \apj, 779, 131 

\bibitem[Horiuchi et al.(2011)]{Horiuchi 2011} Horiuchi, S., Beacom, J.~F., Kochanek, C.~S., et al.\ 2011, \apj, 738, 154 

\bibitem[H. E. S. S. Collaboration et al.(2011)]{Brun et al. 2011} H. E. S. S. Collaboration,\ 2011, Proc. 25th TEXAS Symposium, Relativistic Astrophysics, preprint (arXiv:1104.5003) 

\bibitem[IceCube Collaboration et al.(2014)]{IceCube 2014} IceCube Collaboration et al.\ 2014, \apj, 796, 109 

\bibitem[Kamae et al.(2006)]{Kamae et al. 2006} Kamae, T., Karlsson, N., Mizuno, T., Abe, T., \& Koi, T.\ 2006, \apj, 647, 692

\bibitem[Kang et al.(2013)]{Jones 2013} Kang, H., Jones, T.~W., \& Edmon, P.~P.\ 2013, \apj, 777, 25 

\bibitem[Kelner et al.(2006)]{Kelner et al. 2006} Kelner, S.~R., Aharonian, F.~A., \& Bugayov, V.~V.\ 2006, \prd, 74, 034018

\bibitem[Kennicutt(1998)]{Kennicutt 1998} Kennicutt, R.~C., Jr.\ 1998, \apj, 498, 541 

\bibitem[Krymskii(1977)]{Krymskii1977} Krymskii, G.~F.\ 1977, Akademiia Nauk SSSR Doklady, 234, 1306 

\bibitem[Lacki et al.(2010)]{Lacki et al. 2010} Lacki, B.~C., Thompson, T.~A., \& Quataert, E.\ 2010, \apj, 717, 1 

\bibitem[Lacki \& Thompson(2010)]{Lacki2010} Lacki, B.~C., \& Thompson, T.~A.\ 2010, \apj, 717, 196

\bibitem[Lacki et al.(2011)]{Lacki et al. 2011} Lacki, B.~C., Thompson, T.~A., Quataert, E., Loeb, A., \& Waxman, E.\ 2011, \apj, 734, 107

\bibitem[Lacki et al.(2014)]{Lacki et al. 2014} Lacki, B.~C., Horiuchi, S., \& Beacom, J.~F.\ 2014, \apj, 786, 40 

\bibitem[Lacki \& Thompson(2013)]{Lacki & Thompson 2013} Lacki, B.~C., \& Thompson, T.~A.\ 2013, \apj, 762, 29

\bibitem[Lemoine-Goumard et al.(2012)]{Lemoine-Goumard et al. 2012} Lemoine-Goumard, M., Renaud, M., Vink, J., et al.\ 2012, \aap, 545, A28

\bibitem[Lichti et al.(1978)]{Lichti 1978} Lichti, G.~G., Bignami, G.~F., \& Paul, J.~A.\ 1978, \apss, 56, 403 

\bibitem[Lien \& Fields(2009)]{Lien & Fields 2009} Lien, A., \& Fields, B.~D.\ 2009, \jcap, 1, 47

\bibitem[Loeb \& Waxman(2006)]{Loeb & Waxman 2006} Loeb, A., \& Waxman, E.\ 2006, \jcap, 5, 3

\bibitem[Longair(1981)]{Longair 1981} Longair, M.~S.\ 1981, High energy Astrophysics, 1st edn., Cambridge Univ.
Press, Cambridge and New York, p. 420 

\bibitem[Meneguzzi et al.(1971)]{Mene1971} Meneguzzi, M., Audouze, J., \& Reeves, H.\ 1971, \aap, 15, 337

\bibitem[Meyer(1985)]{Meyer 1985} Meyer, J.-P.\ 1985, \apjs, 57, 173

\bibitem[Miller et al.(2007)]{Miller 2007} Miller, M.~L., Reygers, K., Sanders, S.~J., \& Steinberg, P.\ 2007, Annual Review of Nuclear and Particle Science, 57, 205

\bibitem[Morlino \& Blasi(2016)]{Blasi 2016} Morlino, G., \& Blasi, P.\ 2016, \aap, 589, A7 

\bibitem[Morlino \& Caprioli(2012)]{Morlino & Caprioli 2012} Morlino, G., \& Caprioli, D.\ 2012, \aap, 538, A81

\bibitem[Mori(2009)]{Mori 2009} Mori, M.\ 2009, Astroparticle Physics, 31, 341 

\bibitem[Murase et al.(2013)]{Murase et al. 2013} Murase, K., Ahlers, M., 
\& Lacki, B.~C.\ 2013, \prd, 88, 121301 

\bibitem[Murase \& Waxman(2016)]{Murase2016} Murase, K., \& Waxman, E.\ 2016, \prd, 94, 103006

\bibitem[Nolan et al.(2012)]{Nolan et al. 2012} Nolan, P.~L., Abdo, A.~A., Ackermann, M., et al.\ 2012, \apjs, 199, 31

\bibitem[Ohm(2016)]{Ohm 2016} Ohm, S.\ 2016, Comptes Rendus Physique, 17, 585 

\bibitem[Olive \& Particle Data Group(2014)]{Olive 2014} Olive, K.~A., \& Particle Data Group 2014, Chinese Physics C, 38, 090001

\bibitem[Paglione et al.(1996)]{Paglione et al. 1996} Paglione, T.~A.~D., Marscher, A.~P., Jackson, J.~M., \& Bertsch, D.~L.\ 1996, \apj, 460, 295

\bibitem[Paglione \& Abrahams(2012)]{PA2012} Paglione, T.~A.~D., \& Abrahams, R.~D.\ 2012, \apj, 755, 106

\bibitem[Pavlidou \& Fields(2001)]{Pavlidou & Fields 2001} Pavlidou, V., \& Fields, B.~D.\ 2001, \apj, 558, 63

\bibitem[Peng et al.(2016)]{Peng 2016} Peng, F.-K., Wang, X.-Y., Liu, R.-Y., Tang, Q.-W., \& Wang, J.-F.\ 2016, \apjl, 821, L20

\bibitem[Persic et al.(2008)]{Persic et al. 2008} Persic, M., Rephaeli, Y., \& Arieli, Y.\ 2008, \aap, 486, 143

\bibitem[Planck Collaboration et al.(2016)]{Planck 2016} Planck Collaboration, Ade, P.~A.~R., Aghanim, N., et al.\ 2016, \aap, 594, A13 

\bibitem[Pohl(1993)]{Pohl1993} Pohl, M.\ 1993, \aap, 270, 91

\bibitem[Pohl(1994)]{Pohl1994} Pohl, M.\ 1994, \aap, 287, 453

\bibitem[Rephaeli et al.(2010)]{Rephaeli et al. 2010} Rephaeli, Y., Arieli, Y., \& Persic, M.\ 2010, \mnras, 401, 473

\bibitem[Rephaeli \& Persic(2013)]{Rephaeli2013} Rephaeli, Y., \& Persic, M.\ 2013, Astrophys. Space Sci. Proc., Cosmic Rays in Star-Forming Environments, Vol. 34, Springer-Verlag Berlin, Heidelberg, p. 193

\bibitem[Sakamoto et al.(2008)]{Sakamoto 2008} Sakamoto, K., Wang, J., Wiedner, M.~C., et al.\ 2008, \apj, 684, 957-977 

\bibitem[Salamon \& Stecker(1998)]{Salamon & Stecker 1998} Salamon, M.~H., \& Stecker, F.~W.\ 1998, \apj, 493, 547

\bibitem[Sanders et al.(2003)]{Sanders 2003} Sanders, D.~B., Mazzarella, J.~M., Kim, D.-C., Surace, J.~A., \& Soifer, B.~T.\ 2003, \aj, 126, 1607 

\bibitem[Slane et al.(2014)]{Ellison 2014} Slane, P., Lee, S.-H., Ellison, D.~C., et al.\ 2014, \apj, 783, 33

\bibitem[Stecker(1970)]{Stecker1970} Stecker, F.~W.\ 1970, \apss, 6, 377 

\bibitem[Stecker(1971)]{Stecker 1971} Stecker, F.~W.\ 1971, NASA Special Publication, 249

\bibitem[Stecker \& Venters(2011)]{Stecker & Venters 2011} Stecker, F.~W., \& Venters, T.~M.\ 2011, \apj, 736, 40

\bibitem[Stecker et al.(2012)]{Stecker et al. 2012} Stecker, F.~W., Malkan, M.~A., \& Scully, S.~T.\ 2012, \apj, 761, 128 

\bibitem[Stephens \& Badhwar(1981)]{SB1981} Stephens, S.~A., \& Badhwar, G.~D.\ 1981, \apss, 76, 213

\bibitem[Strong et al.(1976)]{Strong 1976} Strong, A.~W., 
Wolfendale, A.~W., \& Worrall, D.~M.\ 1976, \mnras, 175, 23P 

\bibitem[Strong \& Moskalenko(1998)]{Strong & Moskalenko 1998} Strong, A.~W., \& Moskalenko, I.~V.\ 1998, \apj, 509, 212

\bibitem[Strong et al.(2007)]{strong07} Strong, A.~W., Moskalenko, I.~V., \& Ptuskin, V.~S.\ 2007,
  Annual Review of Nuclear and Particle Science, 57, 285  

\bibitem[Strong et al.(2010)]{Strong et al. 2010} Strong, A.~W., Porter, T.~A., Digel, S.~W., et al.\ 2010, \apjl, 722, L58

\bibitem[Thompson, Quataert and Waxman(2007)]{Thompson2007} Thompson, T.~A., Quataert, E., \& Waxman, E.\ 2007, \apj, 654, 219

\bibitem[Torres(2004)]{Torres 2004} Torres, D.~F.\ 2004, \apj, 617, 966

\bibitem[Torres et al.(2012)]{Torres2012} Torres, D.~F., Cillis, A., Lacki, B., \& Rephaeli, Y.\ 2012, \mnras, 423, 822 

\bibitem[Tully et al.(2009)]{Tully2009} Tully, R.~B., Rizzi, L., Shaya, E.~J., et al.\ 2009, \aj, 138, 323

\bibitem[Tunnard et al.(2015)]{Tunnard 2015} Tunnard, R., Greve, T.~R., Garcia-Burillo, S., et al.\ 2015, \apj, 800, 25 

\bibitem[VERITAS collaboration (2015)]{Fleischhack 2015} VERITAS collaboration, 2015, PoS ICRC, p. 745

\bibitem[Voelk(1989)]{voelk} Voelk, H.~J.\ 1989, \aap, 218, 67 

\bibitem[Wang 
\& Fields(2014)]{Wang & Fields 2014} Wang, X., \& Fields, B.~D.\ 2014, American Institute of Physics Conference Series, 1595, 231

\bibitem[Wang 
\& Fields(2016)]{Wang:2015omn} Wang, X., \& Fields, B.~D.\ 2016, PoS ICRC {\bf 2015}, 905 (2016).

\bibitem[Wik et al.(2014)]{Nustar} Wik, D.~R., Lehmer, B.~D., 
Hornschemeier, A.~E., et al.\ 2014, \apj, 797, 79 

\bibitem[Wilson et al.(2014)]{Wilson 2014} Wilson, C.~D., Rangwala, N., Glenn, J., et al.\ 2014, \apjl, 789, L36

\bibitem[Woosley \& Weaver(1995)]{Woosley & Weaver 1995} Woosley, S.~E., \& Weaver, T.~A.\ 1995, \apjs, 101, 181

\bibitem[Yoast-Hull et al.(2013)]{YH2013} Yoast-Hull, T.~M., Everett, J.~E., Gallagher, J.~S., III, \& Zweibel, E.~G.\ 2013, \apj, 768, 53

\bibitem[Yoast-Hull et al.(2014)]{YH2014} Yoast-Hull, T.~M., Gallagher, J.~S., III, Zweibel, E.~G., \& Everett, J.~E.\ 2014, \apj, 780, 137

\bibitem[Yoast-Hull et al.(2015)]{YH2015} Yoast-Hull, T.~M., Gallagher, J.~S., \& Zweibel, E.~G.\ 2015, \mnras, 453, 222 

\bibitem[Yoast-Hull et al.(2017)]{YH2017} Yoast-Hull, T.~M., Gallagher, J.~S., III, Aalto, S., \& Varenius, E.\ 2017, \mnras, 469, L89

\end{thebibliography}
\end{document}